\documentclass[conference]{IEEEtran}
\IEEEoverridecommandlockouts
\usepackage{packages}

\begin{document}

\title{Techno-economic analysis of PV-battery systems in Switzerland\\
\thanks{This research was carried out within the Nexus-e project and is also part of the activities of the Swiss Centre for Competence in Energy Research on the Future Swiss Electrical Infrastructure (SCCER-FURIES), which is financially supported by the Swiss Innovation Agency (Innosuisse - SCCER program).}
}

\author{
\IEEEauthorblockN{Xuejiao Han, Gabriela Hug}
\IEEEauthorblockA{\textit{Department of Information Technology and Electrical Engineering}\\
\textit{ETH Zürich}\\ 
Zürich, Switzerland\\
\{xuhan, hug\}@eeh.ee.ethz.ch}
\and
\IEEEauthorblockN{Jared Garrison}
\IEEEauthorblockA{\textit{Research Center for Energy Networks}\\
\textit{ETH Zürich}\\
Zürich, Switzerland \\
garrison@fen.ethz.ch}
}

\maketitle

\thispagestyle{plain}
\pagestyle{plain}

\begin{abstract}
This paper presents a techno-economic optimization model to analyze the economic viability of a \gls{pvb} system for different customer groups in Switzerland clustered based on their annual electricity consumption, rooftop size, annual irradiation and location. The simulations for a static investment model are carried out for years 2020-2050 and a comprehensive sensitivity analysis is conducted to investigate the impacts of individual parameters such as costs, load profiles, electricity prices and tariffs, etc.
Results show that while combining \gls{pv} with batteries already results in better net present values than \gls{pv} alone for some customer groups today, the payback periods fluctuate between 2020 and 2035 due to the mixed effects of policy changes, costs and electricity price developments. 
The optimal \gls{pv} and battery sizes increase over time and in 2050 the \gls{pv} investment is mostly limited by the rooftop size.
The economic viability of \gls{pvb} system investments varies between different customer groups and the most attractive investment (i.e., that has the shortest payback period) is mostly accessible to customer groups with higher annual irradiation and electricity demand.
In addition, investment decisions are highly sensitive to payback periods, future costs, electricity prices and tariff developments. 
Lastly, the grid impact of the \gls{pvb} system  deployments are investigated by analyzing the residual Swiss system load profiles. The dynamics of residual load profiles caused by the seasonal, daily and hourly patterns of the solar generation emphasizes the need for flexible resources with fast ramping capabilities. 
\end{abstract}


\section{Introduction}
\subsection{Motivation}
Solar energy is widely recognized as a solution to tackle climate change by lowering worldwide greenhouse gas emissions from the energy sector \cite{web_IPCC}. 
After a slowdown in 2018, the global solar energy market experienced a strong recovery in 2019, reaching 627 GW of cumulative \gls{pv} installations \cite{IEAPVPS2019}. This capacity accounts for nearly 3\% of the global electricity demand and contributes to around a 5\% reduction in worldwide electricity related CO$_2$ emissions \cite{pvps2020snapshot}.

Major drivers for the increasing PV penetration are the provision of subsidies and the overall decreasing costs. 
But subsidies that aim to compensate the capital-intensive \gls{pv} investment are changing: feed-in tariffs are decreasing continuously while injection remunerations that are paid by the local \glspl{dso} (which we will refer to as the injection tariff in the following context) are already or will soon be lower than the retail tariff, which encourages self-consumption of \gls{pv} generation.
One of the means to enable the further development of \gls{pv} installations is the use of battery storage, which is able to increase the \gls{pv} self-consumption rate and also resolve the real-time imbalances caused by forecast errors \cite{han2020distributionally}. 
In the past, high costs and limited combinations of use cases were the greatest barriers for battery installations.
However, as battery prices have declined dramatically over the last decade\footnote{Battery packs decreased from over 1100 \$/kWh in 2010 to around 150 \$/kWh in 2019.}, mainly driven by developments in the \gls{ev} industry, batteries are now considered to be one of the most promising solutions to enable the transition towards renewable energy sources. In addition, with a proper combination of different applications, investments in battery storage units could already be attractive today \cite{stephan2016limiting}.
\subsection{Literature review}

\begin{table*}[!t]
\renewcommand{\arraystretch}{1.1}
\begin{threeparttable}
\centering
\scriptsize
\caption{Overview of existing techno-economic studies of the PVB system.}
\label{tab:literatureReview}
\begin{tabular}{|l|c|c|c|c|c|c|c|}
\hline
\textbf{Ref.} & \textbf{Year} & \textbf{Region} & \textbf{Model type} & \textbf{Battery type} & \textbf{Main economic indicator} & \textbf{Main sensitivity analysis} & \textbf{Load}\\ \hline
\cite{stephan2016limiting} & 2016 & GE & Simulation & Lithium-ion & NPV & Battery application combination & S \\ \hline 
\cite{hoppmann2014economic} & 2014 & GE  & Optimization & Lead-acid & NPV & Wholesale electricity price, retail tariff & R \\ \hline 
\cite{truong2016economics} & 2016 & GE  & Simulation & Lithium-ion & ROI & Battery parameters, electricity price, household size & R \\ \hline 
\cite{merei2016optimization} &  2016 & GE & Simulation & Lithium-ion & Annuity costs & Size and cost of PV and battery & R \\ \hline 
\cite{kaschub2016solar}& 2016 & GE & Simulation & n/a & NPV & Load profile, EV profile & R \\ \hline 
\cite{nyholm2016solar}& 2016 & SE  & Optimization & Lithium-ion & SC, SS & Load profile, battery E-rate & R \\ \hline 
\cite{khalilpour2016technoeconomic} & 2016 & AU & Optimization & Lithium-ion \& lead-acid & NPV & Load profile, PVB system cost, electricity tariff & R\\ \hline 
\cite{quoilin2016quantifying} & 2016 & Europe & Simulation & n/a & LCOE  & load profile, location  & S  \\ \hline 
\cite{camilo2017economic}& 2017 & PT  & Simulation &  Lead-acid & NPV, IRR, PI, DPP, LCOE & Consumption mode & S \\ \hline 
\cite{linssen2017techno} & 2017 & GE & Optimization & Lithium-ion & LCOE & Load profile & R  \\ \hline 
\cite{bertsch2017drives} & 2017 & GE, IE  & Simulation & Lithium-ion  & IRR & Load profile & S \\ \hline 
\cite{zhang2017comparative} & 2017 & SE & Optimization & Hydrogen \& lithium-ion & NPV & Operation strategy, PVB cost & R \\ \hline 
\cite{uddin2017techno}  & 2017 & UK & Simulation & Lithium-ion & Net benefit & Battery degradation costs & R\\ \hline 
\cite{hassan2017optimal} & 2017 & UK & Optimization  & Lithium-ion & Annuity cost & Electricity tariff mode, battery capacity & R \\ \hline 
\cite{vieira2017energy} & 2017 & PT & Simulation & Lithium-ion & Electricity bill, NPV & Battery cost, interest rate & S\\ \hline 
\cite{e2017photovoltaic} & 2017 & BE & Simulation & Lithium-ion & LCOE  & Size of PV and battery, storage price & R\\ \hline 
\cite{akter2017comprehensive} & 2017 & AU & Simulation & n/a & LCOE, NPV, IRR, DPBP, PBP & Size of PV and battery & R\\ \hline 
\cite{barcellona2017economic} & 2017 & CH, UK, IT & Optimization &  Lithium-ion \& lead-acid & NPV & Location, battery technology & S \\ \hline 
\cite{schopfer2018economic} & 2018 & CH  & Optimization & Lithium-ion & NPV & PV and battery parameters and cost & R \\ \hline 
\cite{tervo2018economic} & 2018 & US & Simulation & Lithium-ion & LCOE & PVB cost, subsidy, discount rate, battery efficiency & S\\ \hline 
\cite{litjens2018economic} & 2018 & NE & Simulation & Lithium-ion & PI & Battery operation strategy, PVB system size and cost & R \\ \hline 
\cite{say2019power} & 2019 & AU & Simulation & Lithium-ion & DPPB & Discount rate, feed-in-tariff & R  \\ \hline  
\cite{bai2019economic} & 2019 & CN & Simulation & Reused EV battery & NPV & Battery operation strategy & S \\ \hline 
\cite{koskela2019using} & 2019 & FI  & Simulation & Lithium-ion  & Annuity cost  & Electricity price and tariff mode  &  S \\ \hline 
\cite{lazzeroni2020economic}& 2020 & IT & Simulation & n/a & NPV, LCOE, IRR, DPBP & Consumption level, investment scheme & S\\ \hline
\cite{chaianong2020customer} & 2020 & TH & Simulation & Lithium-ion & NPV, LCOE & Retail tariff, battery cost and size & R \\ \hline 
\end{tabular}
    \begin{tablenotes}
      \scriptsize
      \item Note: Definitions of economics indicators can be found in Section \ref{sec:Method}. Load type R and S stand for real-world and synthetic, respectively.
    \end{tablenotes}
  \end{threeparttable}
\vspace{-0.3cm}
\end{table*}
Techno-economic assessments of \gls{pvb} systems have been extensively researched in recent years, especially in Germany where favorable renewable policies are implemented.
As shown in Table \ref{tab:literatureReview}, the existing techno-economic models can be categorized into optimization and simulation models, depending on whether the capacity of \gls{pv} and battery units are optimization variables or simulated as exogenous parameters.
While most of the existing studies focus on applications of \gls{pvb} systems in residential sectors, some also investigate commercial and industry sectors \cite{merei2016optimization,stephan2016limiting,litjens2018economic,bai2019economic}.

Most of the existing research focuses on lithium-ion or lead-acid batteries, however, recent studies have shown that lithium-ion batteries are more viable, techno-economically, than lead-acid batteries \cite{e2017photovoltaic,dhundhara2018techno} thanks to their recent drastic cost reductions and technology improvements.
Some works also investigate hydrogen-based battery units \cite{zhang2017comparative} as well as reused electric vehicle batteries \cite{bai2019economic}. A review of different stationary electricity storage technologies can be found in \cite{abdon2017techno}.

Concerning battery operation strategies, most works \cite{hoppmann2014economic, schopfer2018economic,tervo2018economic,uddin2017techno,e2017photovoltaic} adopt simple rule-based strategies that aim to maximize the self-consumption rate, i.e. surplus \gls{pv} generation is primarily used to charge the battery while any demand deficit is first satisfied by the stored energy in the battery. Some consider hybrid operation strategies, e.g. \cite{vieira2017energy} applies batteries to peak shaving while \cite{litjens2018economic} investigates uses in frequency reserve provision. In \cite{stephan2016limiting}, the benefits of combining different applications of battery storage units are investigated.
As mentioned in \cite{zhang2017comparative}, simple rule-based strategies might underestimate the economic value of the investment and it is indeed important to adopt appropriate operation strategies in the analysis.

Since the input data and parameters such as costs, load profiles, wholesale and retail electricity prices, and local policies vary widely across published studies, different conclusions concerning the economics of \gls{pvb} systems are drawn.
While references \cite{uddin2017techno} and \cite{akter2017comprehensive}, published in 2017, state that the integration of batteries is not attractive at that time in the UK and Australia, \cite{hoppmann2014economic} and \cite{stephan2016limiting}, published in 2014 and 2016, indicate that it could be profitable for certain \gls{pvb} in Germany. However, a comparatively low battery cost, i.e. 171 \euro/kWh + 172 \euro/kW, is assumed in \cite{hoppmann2014economic}. 
The work in \cite{tervo2018economic} shows that pairing \gls{bess} with \gls{pv} systems can improve the economics and performance of a \gls{pvb} system in the US and \cite{vieira2017energy} identifies that the electricity bill could be reduced by 87\% for the considered residential house in Portugal.
Some studies investigate the break-even price of battery units.
For example, \cite{merei2016optimization,kaschub2016solar,hassan2017optimal,schopfer2018economic,chaianong2020customer}, which were published between 2016 and 2020 and simulated battery costs between 138-400 \euro/kWh, concluded that batteries could be profitable for commercial or residential sectors in Belgium, Germany, the UK, Switzerland and Thailand.
In contrast, the study in \cite{linssen2017techno} estimates that the break-even price of \gls{bess} in Germany ranges from 900 to 1200 \euro/kWh, whereas the work in
\cite{bertsch2017drives} finds that battery costs of 500–600 \euro/kWh may make \gls{pvb} systems generally profitable in Germany even without subsidies.




Based on this literature review, the identified research gaps are as follows:

\begin{itemize}
    \item  Most papers consider one single representative household for the entire country, i.e one single price and one tariff for the \gls{pvb} system, thereby neglecting price differences between different \gls{pv}/battery categories, regional differences within one country, and different trade-offs faced by different household groups. This makes it difficult for policy-makers and regulators to learn from these studies.
    \item Most papers assume a simple rule-based battery operation strategy that aims to maximize the self-consumption rate, which underestimates the value of battery investments by ignoring the multi-applications case (e.g. price arbitrage). 
    \item There is limited discussion about battery C-rates (i.e. the rate to quantify the maximum discharging rate of the battery as a reference to its maximum capacity) and most works only make energy-related cost assumptions.
    \item Specific types of load profiles are utilized for the analyses, e.g. scaled aggregated load profiles as well as synthetic profiles or real measurements taken from individual households. But there is limited analysis of the impact of load profiles, which are expected to affect the \gls{pv} self-consumption rate and the profitability of battery units.
    \item There is almost no analysis of the grid impact (e.g., maximum hourly injection and ramping etc.) of \gls{pvb} system installations.
\end{itemize}
This work aspires to address these gaps and presents a static investment optimization model to assess the economics of \gls{pvb} systems by minimizing the total investment and operational costs over a 30-year horizon. The optimization is conducted for a variety of customer groups in Switzerland in the years from 2020 through 2050. The customer group's heterogeneity is modeled using different rooftop sizes, annual irradiation and electricity consumption values, individual load profiles and geographical regions.

\subsection{Status of PV and BESS in Switzerland}
To support the implementation of the Energy Strategy 2050 \cite{bfe2050} and a smooth transition towards a nuclear phaseout, Switzerland introduced different policies to encourage the deployment of renewables, especially \gls{pv} investments, including: a feed-in tariff, investment subsidy, tax rebates and injection remunerations. \gls{pv} is considered to be the most promising renewable resource in Switzerland due to the high social acceptance and the high deployment potential. The solar installation potential on rooftops and building facades in Switzerland is estimated to be 67 TWh (including 17 TWh from facades) \cite{web_bfe_solarpotential}.
As a result, the annual \gls{pv} deployment increased from 26 MW in 2009 to 327 MW in 2019 \cite{IEAPVPS2019}, reaching a cumulative installed capacity of 2.5 GW and accounting for about 3.3\% of the annual Swiss electricity demand in 2019 (i.e. 2.11 TWh of \gls{pv} toward the 63.4 TWh demand). 
However, to achieve the ambitious net-zero greenhouse gas emissions targets by 2050 and to replace the phasing-out nuclear power, nearly 50 GW of new \gls{pv} installations are required by 2050 according to Swisssolar \cite{web_swisssolar_50GW}, which translates into around 1.6 GW of new installations annually.

According to data published by Swisssolar \cite{web_Swisssolar_marketerhebung}, the battery storage market in Switzerland, although still quite small, has experienced an increase in annual installed capacity in the last few years. In 2018, 14.6 MWh were added, while in 2019 new installations increased to 20.4 MWh (including 20.3 MWh lithium-ion and 0.09 MWh lead-acid batteries), leading to a total battery storage capacity of 50.7 MWh. Additionally, the average system size increased from 9.1 kWh in 2018 to 13.5 kWh in 2019, which is consistent with the increase in the average installed \gls{pv} unit size (from 19.4 kW in 2018 to 22.5 kW in 2019).
In addition, around 15\% of newly installed \gls{pv} systems for single-family houses are equipped with battery storage units.

Based on these trends and developments, this work aims to answer questions such as:
\begin{itemize}
    \item How are the \gls{pvb} system economics affected by different customer groups that are categorized by rooftop sizes, annual electricity consumption and irradiation values, and geographical location of deployment?
    \item How does the optimal size of the \gls{pvb} system change across different customer groups?
    \item What are the expected cumulative investments of the \gls{pvb} system at both the regional and the national levels over the coming years?
    \item How sensitive is the economic viability of the \gls{pvb} system to uncertainties related to e.g. costs, load profiles, electricity prices, etc. and which are the driving factors?
    \item What are the potential challenges and opportunities for investors, retailers, electricity system operators and policy-makers?
\end{itemize}
The rest of the paper is organized as follows: 
Section \ref{sec:Data} describes the data and assumptions in this research.
Mathematical formulations of the proposed optimization model are given in Section \ref{sec:Method}. 
Section \ref{sec:Results} analyzes the results and a further discussion of the results from different perspectives is given in Section \ref{sec:Discussions}. Finally, limitations of this work and conclusions are stated in Section \ref{sec:Limitation and future work} and Section \ref{sec:Conclusions}, respectively.


\section{Data}
\label{sec:Data}





\subsection{General assumptions}
We run the static investment model for the examined years 2020-2050 with a step of 5 years and the lifetime of the \gls{pvb} system is assumed to be the same as the lifetime of \gls{pv}, i.e. 30 years. \Gls{wacc} is set to be 4\% \cite{hoppmann2014economic} and the amortization period is the same as the lifetime of the invested unit. Since the lifetime of battery units are in general shorter than 30 years, a battery replacement is assumed and the potential remaining value of the last reinvested battery by the end of the \gls{pvb} system lifetime is also calculated.

\subsection{Rooftop potential and data clustering}
\label{subsec:Data_B}
We focus on rooftop solar and simulate each potential rooftop based on the Sonnendach dataset \cite{sonnendach_berechnung}, which analyzes the solar generation potential for Switzerland by accounting for the roof area, orientation, tilt, utilization type and region. The high level of detail in this dataset thus enables a high level of granularity in our simulation results.
According to \cite{sonnendach_berechnung}, only buildings with roof areas greater than 10 m$^2$ and an annual solar irradiation higher than 1000 kWh/m$^2$ should be considered. The availability factors of the rooftops, which reduce the effective rooftop area, range between 42\% and 80\% depending on building types, roof sizes and tilt. This range accounts for the possible unavailability of the roof areas due to factors such as obstructions, windows and shadings (for details see page 7 of \cite{sonnendach_berechnung}). After accounting for these factors, the theoretically available rooftop area is reduced from 630 km$^2$ to 304 km$^2$ (i.e. 105 GW to 51 GW assuming 6 m$^2$/kWp).
We further process the data by focusing on detached buildings (i.e. Einzelhaus) with warm water consumption that account for around 94\% of the potential solar generations and exclude potentials from bridges, high buildings, buildings under construction, etc.
Finally, the total potential rooftop area modeled in this work equals 224 km$^2$ (i.e. 37 GW), which corresponds to 3'795'145 rooftop data entries.

To lower the computational burden, these nearly 4 million data entries are clustered into different groups depending on their annual irradiation, roof sizes, warm water consumption (which is used to approximate their electricity consumption), and geographical regions:
\begin{itemize}
    \item \textbf{IRR1}-\textbf{IRR5}: 5 irradiation categories in kWh/m$^2$/year with a step of 150 kWh/m$^2$/year, i.e. 1'000-1'150, 1'150-1'300, 1'300-1'450, 1'450-1'600 and >1'600;
    \item \textbf{A1}-\textbf{A40}: 40 roof size categories with a step of  6 m$^2$ between 12 m$^2$ and 60 m$^2$, a step of  12 m$^2$ between 60 m$^2$ and 180 m$^2$, a step of  30 m$^2$ between 180 m$^2$ and 600 m$^2$, a step of  300 m$^2$ between 300 m$^2$ and 1'200 m$^2$, a step of 600 m$^2$ between 1'200 m$^2$ and 2'400 m$^2$ and a step of 1'200 m$^2$ between 2'400 m$^2$ and 6'000 m$^2$;
    \item \textbf{L1}-\textbf{L11}: 11 annual electricity consumption categories in kWh/year\footnote{Since the annual electricity consumption data is not available, we approximate the annual electricity load as 125\% of the warm water consumption \cite{web_warmwasser_energielexikon,web_warmwasser_VHS,web_energieverbrauch_rwiessen}. The corresponding warm water consumption levels in kWh/year are: 0-1'280, 1'280-2'000, 2'000-2'800, 2'800-3'600, 3'600-4'400, 4'400-6'000, 6'000-10'400, 10'400-20'000, 20'000-24'000, 24'000-120'000 and >120'000.}, i.e. 0-1'600, 1'600-2'500, 2'500-3'500, 3'500-4'500, 4'500-5'500, 5'500-7'500, 7'500-13'000, 13'000-25'000, 25'000-30'000, 30'000-150'000 and >150'000;
    \item \textbf{REG1}-\textbf{REG26}: 26 regions corresponding to the 26 cantons in Switzerland.
\end{itemize}
After clustering, all data entries are categorized into 5*40*11*26 = 57'200 groups which we will refer to as customer groups in the following context. We analyze the economic viability of \gls{pvb} systems across the nearly 4 million rooftops considered in Switzerland by evaluating each customer group using the median values from within each group. 
%
%
\subsection{Parameters of the \gls{pvb} system}
\begin{figure}[t]
  \centering
    \includegraphics[width=.4\textwidth]{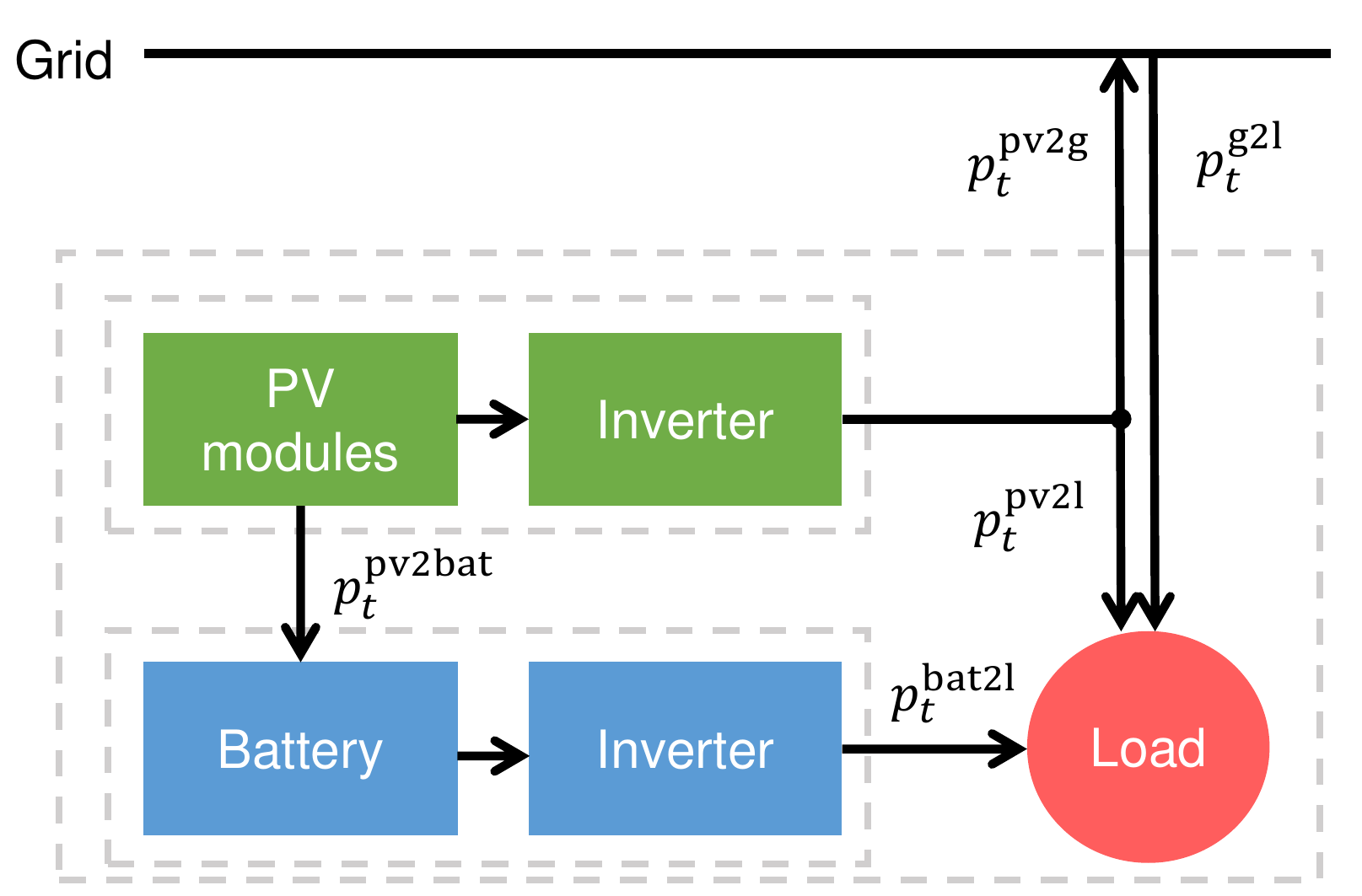}
    \caption{Structure and power flows of the modeled \gls{pvb} system.}
    \label{fig_PVBsys_structure}
\vspace{-0.3cm}
\end{figure}

Each one of the 57'200 customer groups faces an investment optimization problem for a \gls{pvb} system. 
The fundamental model created for the \gls{pvb} system consists of five components: the \gls{pv} module, the battery unit, the hybrid inverter, the load and the grid. The structure of the \gls{pvb} system and the power flows modeled between different components are illustrated in Fig.~\ref{fig_PVBsys_structure}. 
The battery unit is assumed to be AC-coupled since compared to DC-coupling AC-coupling provides higher operational flexibility although it requires an additional battery inverter.

\begin{table}[]
\renewcommand{\arraystretch}{1.2}
\centering
\begin{threeparttable}
\caption{Parameters of the \gls{pvb} system.}
\label{tab:PVBparameter}
\centering
\scriptsize
\begin{tabular}{|l|c|c|c|}
\hline
\textbf{Category} & \textbf{Parameter} & \textbf{Adopted value} & \textbf{Source} \\ \hline
PV & Investment cost & 754\textasciitilde2'786 \euro/kWp & \cite{bauer2017potential} \\ \cline{2-4} 
& Operational cost & 1.7\textasciitilde2.6 cent/kWh & \cite{bauer2017potential} \\ \cline{2-4} 
& Module efficiency & 17\% & \cite{bauer2017potential} \\ \cline{2-4} 
& Inverter efficiency & 98\% & \cite{bauer2017potential} \\ \cline{2-4} 
& Performance ratio & 80\% & \cite{bauer2017potential} \\ \cline{2-4} 
& Lifetime & 30 years & \cite{bauer2017potential} \\ \cline{2-4} 
& Area requirement & 6 m$^2$/kWp & \cite{bauer2017potential} \\ \hline
Battery & Investment cost & \begin{tabular}[c]{@{}c@{}}295\textasciitilde459 \euro/kWh \\ + 249\textasciitilde388 \euro/kW\end{tabular} & \cite{schmidt2019projecting} \\ \cline{2-4} 
 & Operational cost & \begin{tabular}[c]{@{}c@{}} 3.7\textasciitilde5.7 \euro/kW/year \\+ 1.1\textasciitilde1.7 \euro/MWh \end{tabular} & \cite{schmidt2019projecting} \\ \cline{2-4} 
 & Lifetime & 13 years & \cite{schmidt2019projecting} \\ \cline{2-4} 
 & Depth of discharge & 100\% & \cite{schmidt2019projecting} \\ \cline{2-4} 
 & \begin{tabular}[c]{@{}c@{}}Charging/discharging\\ efficiency \end{tabular} & 93\% & \cite{schmidt2019projecting} \\ \cline{2-4} 
& Inverter efficiency & 100\% & n/a \\ \cline{2-4} 
 & Self-discharge & 0\% & \cite{schmidt2019projecting} \\  \hline
PVB system & Degradation rate & 0.5\% per year & \cite{bauer2017potential,linssen2017techno} \\ \hline 
\end{tabular}
    \begin{tablenotes}
      \scriptsize
      \item Note: Original values are converted to Euros based on the exchange rate of 0.91 EUR/CHF and 0.85 EUR/USD.
    \end{tablenotes}
  \end{threeparttable}
\vspace{-0.4cm}
\end{table}

Table \ref{tab:PVBparameter} gives the parameters  for the considered \gls{pvb} system using 2020 as the reference year. 
Based on historical \gls{pv} installation data of Switzerland \cite{web_Swisssolar_marketerhebung}, although the average installed capacity of \gls{pv} is increasing, most of the recent investments are still small-scale. For example, \gls{pv} categories smaller than 1000 kWp account for almost all \gls{pv} deployments in 2019, i.e. <30 kWp (40\%), 30-100 kWp (16\%) and 100-1000 kWp (39\%), while >1000 kWp \gls{pv} investments make up the remaining 5\% of the total installed capacity.
Therefore, we include five \gls{pv} categories (i.e., 0-6 kWp, 6-10 kWp, 10-30 kWp, 30-100 kWp, >100 kWp) and limit the minimum and maximum capacity to 2 kWp and 50 MWp, which covers most of the potential investments and also corresponds to the range of \gls{pv} units that could apply for the one-time investment subsidies in Switzerland \cite{bferegulation}.
The most commonly used batteries combined with a small-scale \gls{pv} are lithium-ion and lead-acid batteries. Although lead-acid batteries have lower capital costs, lithium-ion batteries are proven to be more cost-efficient as a result of better \gls{dod} and cycle life \cite{barcellona2017economic,gupta2020levelized}. In addition, the Swiss battery market is dominated by lithium-ion with only a negligible amount of lead-acid batteries installed in recent years; we therefore only consider lithium-ion batteries in this work.
Note that the costs shown in Table~\ref{tab:PVBparameter} are for the year 2020 and are given as ranges since they vary according to the invested unit size and the considered scenario. 
As the assumed battery costs vary greatly between different studies, ranging from 250 \euro/kWh to 1883 \$/kWh, we provide, for comparative reasons, a list of the cost assumptions made by some recent works in Table \ref{tab:literatureReview_batterycost} along with the cost data selected in our simulations which are based on \cite{schmidt2019projecting}.
Future investment and operational costs for \gls{pv} and batteries are estimated using projections from \cite{bauer2017potential} and \cite{lebedeva2018li}\footnote{Data for missing years are estimated using an interpolation or extrapolation method.}, respectively. Details of the costs for future years are provided in Appendix~\ref{Apdx:App-futurecosts_pv} and Appendix~\ref{Apdx:App-futurecosts_battery}.
\begin{table*}[htbp]
\renewcommand{\arraystretch}{1.1}
\centering
\tiny
\caption{Overview of lithium-ion battery system costs.}
\label{tab:literatureReview_batterycost}
\begin{tabular}{|l|c|c|c|c|c|c|c|}
\hline
\textbf{Ref.} & \textbf{Year} & \textbf{Country} & \textbf{Battery specifics} & \multicolumn{3}{c|}{\textbf{Cost}} & \textbf{Future development} \\ \hline
&  & & & \textbf{Energy-related} & \textbf{Power-related} & \textbf{Others} & \\ \hline
\cite{merei2016optimization} & 2016 & GE & n/a & 2015: 1000 \euro/kWh & n/a & n/a& 2035: 375 \euro/kWh \\ \hline
\cite{kaschub2016solar} & 2016 & GE & n/a & 2018: 500 \euro/kWh & n/a & OM cost: 1\% of investment cost & n/a \\ \hline
\cite{bertsch2017drives} & 2016 & GE & 0-100 kWh & 500 \euro/kWh  & n/a & Installation cost: 1330 \euro & n/a\\ \hline
\cite{nykvist2015rapidly} & 2015 & US & BEV & 2014: 300 \$/kWh & n/a& n/a& Learning rate: 6\textasciitilde9\% \\ \hline
\cite{cole2016utility} & 2016 & US & 8-hour battery, utility-scale & 2015: 500 \$/kWh & 4000 \$/kW & n/a&  2015-2050: 34\%, 57\% and 81\% reduction  scenarios \\ \hline
\cite{ardani2016installed} & 2016 & US & 3kW/6kWh, DC-coupled & 500 \$/kWh for battery & 600 \$/kW for inverter & n/a & n/a \\ \hline
\cite{linssen2017techno} & 2017 & GE & n/a & 1000 \euro/kWh & n/a & n/a & n/a \\ \hline
\cite{hassan2017optimal} & 2017 & UK & n/a & 990 \$/kWh & n/a & n/a & n/a \\ \hline
\cite{vieira2017energy} & 2017 & PT & 10.2 kWh & 550 \euro/kWh (480 \euro/kWh for battery)& n/a & n/a & n/a \\  \hline
\cite{e2017photovoltaic} & 2017 & BE & 0.5 kW/kWh & 600 \euro/kWh & 500 \euro/kW for inverter & Installation cost: 200 \euro & n/a \\ \hline 
\cite{akter2017comprehensive} & 2017 & AU & 4-12 kWh & 300 AUD/kWh & \begin{tabular}[c]{@{}c@{}}700 AUD/kW \\+ 400 AUD/kW for inverter\end{tabular} & n/a & n/a \\ \hline
\cite{barcellona2017economic} & 2017 & CH,UK,IT & 128 Wh/kg & 320 \euro/kWh & n/a & \begin{tabular}[c]{@{}c@{}}Installation cost: 100 \euro +2\euro/kg  \\ Inverter cost: 800 \euro \end{tabular}& n/a \\ \hline
\cite{abdon2017techno} & 2017 & n/a & 1 MW, NCA/LTO & 923 \euro/kWh  & 162 \euro/kW &n/a & n/a\\\hline
\cite{curry2017lithium} & 2017 & n/a & n/a & 2016: 273 \$/kWh for battery & n/a & n/a & Learning rate: 19\% \\ \hline
\cite{ralon2017electricity} & 2017 & n/a & n/a & 2016: 200\textasciitilde840 \$/kWh & n/a & n/a & 2030: 145\textasciitilde480 \$/kWh \\ \hline
\cite{schmidt2017thefuture} & 2017 & GE & 0.33 C-rate & 2016: 1883 \$/kWh & n/a & OM cost: 0 & 2030: 524 \$/kWh; 2040: 397 \$/kWh \\ \hline
\cite{schopfer2018economic} & 2018 & CH & n/a &250\textasciitilde1000 \euro/kWh & n/a & \begin{tabular}[c]{@{}c@{}}OM cost: 1\% of inv. cost  \\ Replacement cost: 50\% of inv. cost \end{tabular}& n/a \\ \hline
\cite{tervo2018economic} & 2018 & n/a & 14 kWh Powerwall & 2017: 393 \$/kWh & n/a & \begin{tabular}[c]{@{}c@{}}BOS cost: 700 \$ \\ Installation cost: 1000 \$  \end{tabular}& \begin{tabular}[c]{@{}c@{}}In 15 years: battery cost -50\%, \\ installation cost -25\%, inverter cost 0.12 \$/W \end{tabular}\\ \hline
\cite{litjens2018economic} & 2018 & NE & 25-year lifetime & 200 \euro/kWh & \begin{tabular}[c]{@{}c@{}} 150 \euro/kW BOS \\+ 150 \euro/kW EPC \end{tabular} & OM cost: 1\% of inv. cost  & n/a \\ \hline
\cite{few2018prospective} & 2018 & n/a & n/a & 2020: 165\textasciitilde548 \$/kWh for battery & n/a &n/a  & 2030: 120\textasciitilde250 \$/kWh \\ \hline
\cite{web_lazards_LCOE} & 2018 & US & 10kW/40kWh & 639\textasciitilde780 \$/kWh & 130\textasciitilde174 \$/kW & OM cost: 1.79\textasciitilde2.2\% of inv. cost  & n/a \\ \hline
\cite{say2019power} & 2019 & AU & n/a& 900 AUD/kWh & n/a & n/a & -8\%/year \\ \hline
\cite{koskela2019using} & 2019 & FN & n/a & 2020: 100\textasciitilde200 \euro/kWh & 80\textasciitilde110 \euro/kW & Installation cost: 200\textasciitilde400 \euro/kWh & n/a\\\hline
\cite{schmidt2019projecting} & 2019 & n/a &n/a  & 2015: 802 \$/kWh & 678 \$/kW & OM cost: 10 \$/kW-year + 3 \$/MWh & \begin{tabular}[c]{@{}c@{}} Using 2015 cost as reference, costs for 2020-2050 \\ decrease to 55\%, 34\%, 23\%, 18\%,
16\%, 15\%, 14\%  \end{tabular}\\ \hline
\cite{mongird2019energy} & 2019 & n/a & 1kW-100MW & 2018: 271 \$/kWh & 388 \$/kW for BOP&
\begin{tabular}[c]{@{}c@{}}OM cost: 10 \$/kW-year + 0.3 \$/MWh \\ Installation cost: 101 \$/kWh \end{tabular}&  
\begin{tabular}[c]{@{}c@{}}2025: 306  \$/kW + 189 \$/kWh capital cost \\ + 96 \$/kWh installation cost\\ OM cost: 8 \$/kW-year + 0.3 \$/MWh \end{tabular}\\ \hline
\cite{chaianong2020customer} & 2020 & TH & 6.5 kWh/kW, AC-coupled & 500\textasciitilde1000 \$/kWh & n/a& n/a& -4\%/year to -12\%/year \\ \hline
\end{tabular}
\vspace{-0.2cm}
\end{table*} 
\subsection{Load and generation profiles}
We use synthetic load profiles for individual households generated using the ''LoadProfileGenerator'' \cite{pflugradt2017load} with the location set as Munich.
Then for each customer group, the load profiles are scaled so that the total consumption matches the annual electricity demand approximated using the warm water consumption.
To model the load profile of different consumption categories (i.e. L1-L11), we use different predefined household settings of ''LoadProfileGenerator'' detailed as follows:
\begin{itemize}
    \item L1: predefined household CHR07 (i.e. single, employed) with an annual electricity consumption of 1'502 kWh;
    \item L2: predefined household CHR02 (i.e. couple, 30-64 age, both employed) in energy saving mode with an annual electricity consumption of 1'864 kWh;
    \item L3: predefined household CHR02 (i.e. couple, 30-64 age, both employed) in energy intensive mode with an annual electricity consumption of 3'346 kWh;
    \item L4: predefined household CHR04 (i.e. couple, 30-64 age, 1 employed, 1 at home) with an annual electricity consumption of 4'677 kWh;
    \item L5: predefined household CHR03 (i.e. family, 1 child, both employed) with an annual electricity consumption of 5'460 kWh;
    \item L6: predefined household CHR05 (i.e. family, 3 children, both employed) with an annual electricity consumption of 6'689 kWh;
    \item L7-L11: a combination of predefined household CHR02 and CHR03 with an annual electricity consumption of 8'826 kWh. The electricity consumption of buildings with multiple households are assumed to fall into these consumption categories.
\end{itemize}
 
Solar irradiation profiles are based on historical hourly data from MeteoSwiss \cite{web_meteoswiss}, using data of stations located in the capital or the main city to represent the profile of each canton. The irradiation profiles are then scaled according to the annual irradiation category collected from the Sonnendach data.
A perfect forecast of \gls{pv} generation is assumed and the generation profile is calculated as the production resulting from the invested module area, module efficiency, inverter efficiency, performance ratio and the irradiation profile (a summary of the \gls{pvb} system parameter inputs used in this work is given in Table \ref{tab:PVBparameter}).
\subsection{Policies and regulations}
To account for the impacts of the legislative and regulatory framework on the investment decisions for \gls{pv} units, we consider available subsidies,  \gls{dso} injection tariffs and tax rebates:
\begin{itemize}
    \item Subsidies: 
    Currently, both an output-based feed-in-tariff subsidy scheme and a capacity-based investment subsidy scheme exist in Switzerland.
    However, the feed-in-tariff scheme is expected to expire in 2022 and due to the long waiting list, only \gls{pv} units registered before July 2012 could qualify to benefit from it \cite{pronovo}. From 2020 on, units above 100 kWp within the feed-in-tariff scheme are obliged to participate in direct marketing that aims to replace the fixed tariff with a more market-oriented remuneration tariff \cite{web_Swisssolar_direktvermarktung}. 
    Units ranging from 2 kWp to 50 MWp can apply for the one-time investment subsidy that could cover up to 30\% of their investment costs based on the installed capacity and the \gls{pv} category \cite{bferegulation}. The current one-time investment subsidy is valid until 2030, but recent reports indicate that the Swiss federal council is planning a possible extension to 2035 \cite{web_pvmagazine_renewsupport}. 
    \item \gls{dso} injection tariffs:
    To account for income earned from \gls{pv} generation that is fed back into the local electricity grid, we include the injection tariffs that are set by regional \glspl{dso}. Since these injection tariffs vary from \gls{dso} to \gls{dso}, we use data available from~\cite{pvtarif} and make an estimation of the average value for each canton as \gls{dso} regions and cantons are only partially congruent.
    The inclusion of this injection tariff is important for quantifying the revenue earned from \gls{pv} generation that is not self consumed. Even more critically, it is needed to quantify the economic benefits of the \gls{pv}-batteries that help increase the earnings of the \gls{pvb} system by reducing the \gls{pv} generation sold at this injection tariff by storing for later use as self consumption. Sensitivity analysis is conducted to analyze the impact of injection tariffs.
    \item Tax rebates:
     The available tax rebate covers 20\% of the net investment costs (i.e., investment cost minus the investment subsidy) in all Swiss cantons~\cite{energieschweiz}.
     We assume these tax rebates to remain constant until 2050.
\end{itemize}
Policies and regulations modeled in the Baseline scenario including tariffs and the \gls{wacc} assumption are summarized in Table \ref{tab:Policyparameter}. While the investment subsidy and \gls{dso} injection tariffs are based on the current year's information (i.e. 2020), we assume the retail and wholesale electricity prices for 2020 using the historical 2018 data from \cite{web_elcom} and \cite{web_EPEX}, respectively. In the Baseline scenario, consumers are assumed to have no access to the hourly wholesale market and the electricity injected back into the grid is reimbursed at the regional injection tariff.
The regional injection tariff is assumed to decrease 10\% per year. However, if the injection tariff in a given year and in a given region drops below the Swiss average annual wholesale price of that year, the \gls{pv} injection in that region is instead paid at that average annual wholesale price.
This assumption is based on the guidelines provided in the Swiss Energy Ordinance \cite{EnV2017} that requires the remuneration to be based on the costs incurred by the grid operator for the purchase of equivalent electricity from third parties or its own production facilities.
Details of the regional injection tariff can be found in Appendix~\ref{Apdx:App-PVinjectiontariff}.
\begin{table}[t]
\renewcommand{\arraystretch}{1.1}
\centering
\begin{threeparttable}
\caption{Input parameters for modeled policies and regulations.}
\label{tab:Policyparameter}
\scriptsize
\begin{tabular}{|l|c|c|}
\hline
\textbf{Parameter} & \textbf{Value} & \textbf{Source} \\ \hline
Investment subsidy & 909 \euro\; + 273\textasciitilde309 \euro/kW  & \cite{bferegulation} \\ \hline
Investment subsidy change & -2\%/year & n/a \\ \hline
Investment subsidy expires & 2030 & \cite{bferegulation}\\ \hline
DSO injection tariff & 5.7\textasciitilde 11.8 cent/kWh & \cite{pvtarif}  \\ \hline
DSO injection tariff change & up to -10\%/year\footnote{Detailed yearly development of the injection tariff also depends on the average wholesale market price assumption of the corresponding year.} & \cite{schopfer2018economic} \\  \hline
Retail el. tariff & 12.3\textasciitilde35.4 cent/kWh & \cite{web_elcom} \\  \hline
Retail el. tariff change & +1\%/year & \cite{hoppmann2014economic} \\  \hline
Wholesale el. tariff & 0\textasciitilde 161.4 \euro/MWh & \cite{web_EPEX} \\  \hline
Wholesale el. tariff change & +1.5\%/year & \cite{hoppmann2014economic} \\ \hline
Tax rebate & 20\% of net investment cost & \cite{energieschweiz} \\ \hline
WACC & 4\% & \cite{hoppmann2014economic} \\ \hline
\end{tabular}
    \begin{tablenotes}
      \scriptsize
      \item Note: the exchange rate is assumed to be 0.91 EUR/CHF.
      \end{tablenotes}
\end{threeparttable}
\vspace{-0.5cm}
\end{table}
\subsection{Scenarios}
The profitability of \gls{pvb} system investments is subject to uncertainties as the future development of \gls{pv} and battery costs, injection tariffs, retail and wholesale market prices, subsidy policies etc. are unknown.
Additionally, in our model, financial parameters such as \gls{wacc} and amortization periods are simplified as a constant value for all modeled \gls{pv} categories, which is likely not the case in reality\footnote{In fact, different potential investors, from individual homeowners to larger industrial operators, might have different needs regarding their desired payback periods as well as different considerations about financing an investment in \gls{pv} including the amount of debt they take on and the interest rate set by their lenders. Additionally, the constant assumptions ignore that some investors have non-economic desires, such as early adopters and innovators who might be driven by environmental issues versus laggards and late majority who might have a higher risk aversion.}.
To investigate how the profitability of \gls{pvb} systems, and consequently the investment decisions, are affected by our assumptions, we conduct a set of one-at-a-time sensitivity analyses on some main parameters, such as the projections of \gls{pv} and battery costs, load profiles, retail and wholesale electricity price developments, \gls{pv} injection tariffs, and the \gls{wacc}. 
Note that the sensitivity scenarios described below are only simulated for the example of the canton of Zurich in 2050, while the Baseline scenario is simulated for 2020-2050 for all cantons. 
\subsubsection{PV and battery cost scenarios}
In addition to the Baseline scenario (as introduced in Table~\ref{tab:PVBparameter}), two additional cost sensitivity scenarios, namely a high cost scenario SC1, and a low cost scenario SC2 are simulated. On average, the high (low) cost scenario corresponds to 15\% higher (lower) costs for the \gls{pv} and 54\% higher (lower) costs for the battery than the Baseline scenario. The differences among the three scenarios vary across the years. The different size categories and details of the cost projections for these three scenarios based on \cite{bauer2017potential} and \cite{schmidt2019projecting} can be found in Appendices~\ref{Apdx:App-futurecosts_pv} and~\ref{Apdx:App-futurecosts_battery}.

\subsubsection{Load profile scenarios}
\begin{figure}[t]
  \centering
    \includegraphics[width=0.52\textwidth]{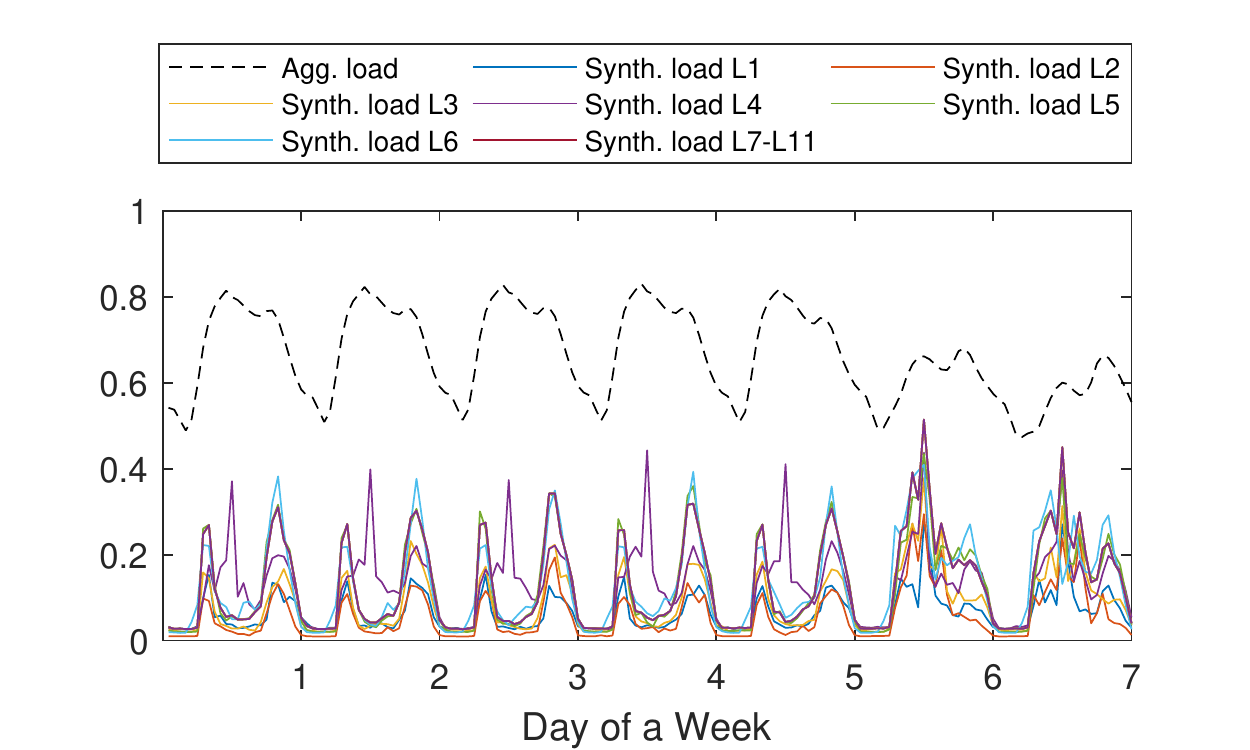}
    \caption{Normalized aggregated load profile for canton Zurich and synthetic individual load profiles for L1-L11.}
    \label{fig_loadprofiles}
\end{figure}

In the Baseline scenario, we model the load profile of consumption categories L1-L11 using different load profiles generated by ''LoadProfileGenerator''.
The work in \cite{linssen2017techno} indicates that using aggregated load profiles leads to higher shares of self-consumption compared to the use of an individual profile. 
Figure \ref{fig_loadprofiles} shows the average weekly normalized aggregated load profile for the canton of Zurich in 2018 together with eleven normalized synthetic load profiles adopted for consumption categories L1-L11. 
It can be seen that the individual load profiles are quite different than the aggregated profile. The individual profiles tend to peak once in the morning and once during the evening while the aggregated profile peaks just once during the day. Furthermore, the aggregated load profile follows a pattern with lower consumption during the weekend whereas the individual customers consume more during the weekend. Such differences could result in different estimates of \gls{pv} self-consumption and evaluations of the battery installations if aggregated load profiles are used instead of individual profiles.
Therefore, we simulate a sensitivity scenario SL, where the synthetic load profiles of all consumption categories are replaced by the corresponding aggregated cantonal load profile, to analyze the impact of using the aggregated load profile.
\subsubsection{Electricity price scenarios}
In the Baseline scenario, we assume that the retail electricity price increases by 1\% per year and the prosumers have no access to the hourly wholesale market. All excess generation injected back into the grid is reimbursed by the regional injection tariff.
The regional injection tariff is assumed to decrease 10\% per year until it reaches the corresponding yearly average Swiss wholesale electricity price, which is assumed to increase by 1.5\% per year. In all years afterwards, the regional injection tariff is instead set equal to the yearly average Swiss wholesale price (see Appendix~\ref{Apdx:App-PVinjectiontariff} for details of the regional injection tariff).

However, it is highly uncertain how the injection tariffs as well as the retail and wholesale electricity prices evolve in future years and it is also unclear to what extent small prosumers will have access to the wholesale market.
To analyze the impact of replacing the injection tariff with the wholesale market price (i.e., simulate the case when end consumers have access to the wholesale market), nine electricity price sensitivity scenarios SP1-SP9 detailed in Table~\ref{tab:SP1_SP9} are simulated similar to the electricity price scenarios modeled in \cite{hoppmann2014economic}.
\begin{table}[t]
\renewcommand{\arraystretch}{1.1}
\centering
\begin{threeparttable}
\caption{Parameters of price scenarios SP1-SP9.}
\label{tab:SP1_SP9}
\begin{tabular}{|l|c|c|}
\hline
\textbf{Scen.} & \textbf{Retail price change} & \textbf{Wholesale price change} \\ \hline
SP1 & \multirow{3}{*}{+0\%/year} & -1\%/year \\ \cline{1-1} \cline{3-3} 
SP2 &  & +1.5\%/year\textbf{*} \\ \cline{1-1} \cline{3-3} 
SP3 &  & +3\%/year \\ \hline
SP4 & \multirow{3}{*}{+1\%/year\textbf{*}} & -1\%/year \\ \cline{1-1} \cline{3-3} 
SP5 &  & +1.5\%/year\textbf{*} \\ \cline{1-1} \cline{3-3} 
SP6 &  & +3\%/year \\ \hline
SP7 & \multirow{3}{*}{+2\%/year} & -1\%/year \\ \cline{1-1} \cline{3-3} 
SP8 &  & +1.5\%/year\textbf{*} \\ \cline{1-1} \cline{3-3} 
SP9 &  & +3\%/year \\ \hline
\end{tabular}
    \begin{tablenotes}
      \scriptsize
      \item Note: values that are the same as the Baseline scenario are noted with an asterisk (\textbf{*}).
      \end{tablenotes}
\end{threeparttable}
\vspace{-0.5cm}
\end{table}
One other sensitivity scenario (i.e. SP10) is simulated to analyze the extreme case of having an injection tariff equal to zero and no access to the hourly wholesale market while the retail prices increase by 1\% per year (same as the Baseline scenario).

\subsubsection{Battery scenario}
In the Baseline scenario, we set the battery costs based on \cite{schmidt2019projecting}, which projects the development of the battery costs using a number of international reports.

Since the battery costs (especially the labor cost) in Switzerland are generally higher than the global average, we create a sensitivity scenario (i.e., SB1) in which we adjust the battery investment cost assumption for 2020 using the current Tesla Powerwall 2 price in Switzerland (i.e. 14'700 CHF equivalent to 13'364 EUR accounting for the total costs incurred for installing a 13.5 kWh Tesla Powerwall 2), while the cost reduction rate over the years remains the same as the Baseline scenario.
Furthermore, we also simulate a sensitivity scenario without any batteries (i.e., SB2) to analyze the financial benefit of installing batteries.

\subsubsection{WACC scenarios}
Cost of capital is defined as the expected rate of return that market participants require in order to attract funds for a particular investment \cite{grabowski2014cost}. In the Baseline scenario, we assume a 4\% \gls{wacc} for all \gls{pvb} system investments.

The value of \gls{wacc} varies over time and between different technologies, e.g. smaller \gls{pvb} systems are mainly invested by households, who face lower \gls{wacc} than investors of larger-sized \gls{pvb} systems. Therefore, we simulate two sensitivity scenarios assuming a 2\% (i.e. SW1) and a 8\% (i.e. SW2) \gls{wacc} to compare against the Baseline assumption.

Table~\ref{tab:scenario_summary} summarizes the main parameter changes of the different sensitivity scenarios compared to the Baseline scenario.
\begin{table}[t]
\renewcommand{\arraystretch}{1.1}
\centering
\scriptsize
\caption{Summary of the sensitivity scenarios.}
\label{tab:scenario_summary}
\begin{tabular}{|l|c|c|}
\hline
\textbf{Scen. name} & \textbf{Changed parameters} & \textbf{Remarks} \\ \hline
SC1-2  & PV and battery costs & \begin{tabular}[c]{@{}c@{}}A high cost (SC1) and \\ a low cost scenario (SC2)\end{tabular}\\ \hline
SL  & Load profile & \begin{tabular}[c]{@{}c@{}}Individual load  profiles replaced \\ by the aggregate profile\end{tabular} \\ \hline
SP1-9 & \begin{tabular}[c]{@{}c@{}}Retail and wholesale \\ el. price development\end{tabular} & \begin{tabular}[c]{@{}c@{}}Access to wholesale market;\\ injection tariff replaced\\ by hourly wholesale price \end{tabular} \\ \cline{1-1} \cline{2-3} 
SP10 &  \begin{tabular}[c]{@{}c@{}}Retail el. price\\ development;\\ injection tariff\end{tabular} & \begin{tabular}[c]{@{}c@{}}No access to wholesale market;\\ injection tariff is zero\end{tabular} \\ \hline
SB1-2  & \begin{tabular}[c]{@{}c@{}}Battery price;\\ battery investment\end{tabular} & \begin{tabular}[c]{@{}c@{}}Battery price adjusted using current\\ Tesla price in Switzerland (SB1);\\battery forced not to be installed (SB2)\end{tabular}\\ \hline
SW1-2  & WACC & \begin{tabular}[c]{@{}c@{}}A 2\% WACC (SW1) and \\ a 8\% WACC scenario (SW2)\end{tabular} \\ \hline
\end{tabular}
\vspace{-0.2cm}
\end{table}

\section{Method}\label{sec:Method}
In this section, the mathematical formulation of the optimization problem is first described, followed by the definitions of the technical and economic indicators used for evaluating the investment decisions. 

In this work, the investment decisions are optimized using a static model. More specifically, for each region and each examined year we run the optimization considering a 30-year lifetime of the \gls{pvb} system.
The simulation optimizes the investment decisions over the full 30-year lifetime by optimizing the operational decisions for all 8760 hours of the examined year and assuming identical operations along with projections for other parameters (e.g., wholesale price and injection tariff) over the remaining lifetime of the \gls{pvb} system, i.e. 29 years.
The model formulation, described below, is applied to each region, hence the region index is omitted in the following equations for simplification.
To optimize the investment and operational decisions, three groups of constraints are considered: 1) investment constraints, 2) operational constraints and 3) system power balance constraints. 
The objective is to minimize the total investment and operating costs of the \gls{pvb} system, which consists of the \gls{pv} unit, the battery unit and the load, over the 30-year simulation horizon. Details of the objective function are given after the constraints are described.
\vspace{-0.2cm}
\subsection{Investment constraints}
Each rooftop in each region $reg\in REG$ is categorized by which customer group $c$ it fits into, defined by the combination of irradiation category $i \in I$, electricity consumption category $j \in J$, and roof size category $k \in K$ (i.e, for each region this is 1 out of 2200 possible customer groups). In other words, the customer group set $C$ with $c \in C$ includes all combinations of irradiation, electricity consumption and roof size categories, i.e. $C=\{(i,j,k): i\in I, j\in J, k\in K\}$. Each combination $(i,j,k)$ is represented by a specific customer group $c$. 

As mentioned, five \gls{pv} candidate units corresponding to five size categories (i.e., 0-6 kWp, 6-10 kWp, 10-30 kWp, 30-100 kWp, >100 kWp) are considered in this work.
Let $P$ denote the set of all these five candidate \gls{pv} categories. 
For each customer group $c$, the sum of the installed capacity $cap^\text{pv}_{p,c}$ over all \gls{pv} categories $p\in P$ is non-negative and limited by the maximum deployment potential $dep_{c}^\text{pv,max}$, which is equal to the corresponding available rooftop area of the customer group divided by the rooftop area required for 1 kWp of \gls{pv} (i.e. 6m$^2$/kWp, provided in Table~\ref{tab:PVBparameter}). Consequently,
\begin{align}
& 0 \leq \sum_p cap^\text{pv}_{p,c}  \leq dep_{c}^\text{pv,max} \label{eq_start}
\end{align} 
The installed capacity $cap^\text{pv}_{p,c}$ of each \gls{pv} category should be greater or equal to the minimum size requirement of that category $cap^\text{pv,min}_p$, i.e.,
\begin{align}
& cap^\text{pv}_{p,c} = u^\text{pv}_{p,c} x^\text{pv}_{p,c} \\
& cap^\text{pv}_{p,c}  \geq u^\text{pv}_{p,c}cap^\text{pv,min}_{p} \label{subeq:inv_min}
\end{align}
where $u^\text{pv}$ is a binary variable that indicates whether the \gls{pv} unit is invested or not and $x^\text{pv}$ is the continuous investment capacity variable. 
All investment decisions are non-negative:
\begin{align}
& cap^\text{pv}_{p,c},\; cap^\text{bat-e}_{c},\; cap^\text{bat-p}_{c} \geq 0 \label{subeq:inv_nonneg}
\end{align}
where $cap^\text{bat-e}$ and $cap^\text{bat-p}$ are the invested energy and power capacity of the PV-battery unit, respectively. Note that the battery C-rate is not fixed and is decided by the invested energy and power capacity of the battery.

\subsection{Operational constraints}
The \gls{pv} generation output $gen^\text{pv}_{t,p,c}$ of \gls{pv} unit $p$ and customer group $c$ at time $t$ is limited by the invested module area $A^\text{pv}$ multiplied by the module efficiency $\eta^\text{pv-mod}$, inverter efficiency $\eta^\text{inv}$, performance ratio $\eta^\text{pv-pf}$ and the solar irradiation $I^\text{pv}$ at time $t$, i.e.,
\begin{align}
& 0  \le gen^\text{pv}_{t,p,c} \le A^\text{pv}_{p,c}\eta^\text{pv-mod}_{p}\eta^\text{inv}_{p}\eta^\text{pv-pf}_{p}I^\text{pv}_{t,c} \label{eq:pv_gen}
\\
& A^\text{pv}_{p,c} = cap^\text{pv}_{p,c}a^\text{pv}_{p}
\end{align}
where $a^\text{pv}_{p}$ is the rooftop area required by each kWp of the installed \gls{pv}. The inequality in constraint (\ref{eq:pv_gen}) allows the possibility of \gls{pv} curtailment.

The \gls{pv}-battery has no direct connection to the grid and, in general, it charges (discharges) when the demand of the customer is lower (higher) than the \gls{pv} generation.
The stored energy of the \gls{pv}-battery unit $E^\text{bat}$ is limited by its maximum \gls{dod} indicated by $DOD^\text{max}$ and the installed energy capacity $cap^\text{bat-e}$:
\begin{align}
& (1-DOD^\text{max}) cap^\text{bat-e}_{c} \le E^\text{bat}_{t,c} \le cap^\text{bat-e}_{c} \label{eq:bat_energy_limit}
\end{align}

The \gls{pv}-battery inflow $p^\text{ch}$ and outflow $p^\text{dis}$ are non-negative and limited by the installed power capacity of the battery $cap^\text{bat-p}$. Mathematically,
\begin{align}
& 0  \le p^\text{ch}_{t,c} \le cap^\text{bat-p}_{c} \\
& 0  \le p^\text{dis}_{t,c} \le cap^\text{bat-p}_{c} 
\end{align}

Finally, the relationship of the storage level $E^\text{bat}$ at the end of each time step across two consecutive time steps is defined by: 
\begin{align}
E^\text{bat}_{t,c}  & = E^\text{bat}_{t-1,c} + \eta^\text{bat,c} p^\text{ch}_{t,c}\Delta_t -  p^\text{dis}_{t,c}\Delta_t/(\eta^\text{bat,d}\eta^\text{bat,inv})  
\end{align}
where $\eta^\text{bat,c}$ and $\eta^\text{bat,d}$ are the charging and discharging efficiencies of the battery, The battery inverter efficiency is denoted as $\eta^\text{bat,inv}$ 
and $\Delta_t$ is the length of one time step.

\subsection{Power balance constraints}
As shown in Fig.\ref{fig_PVBsys_structure}, the power from the \gls{pv} units could be used to 1) charge the battery with $p^\text{pv2bat}$, 2) supply (at least part of) the demand with $p^\text{pv2l}$ or 3) be injected into the grid with $p^\text{pv2g}$.
Note that each customer group $c$ has the choice to invest in any category and any number of \gls{pv} panels as long as the corresponding rooftop size allows.
At each time step, the sum of the power outflows of all \gls{pv} units installed by customer group $c$ should not be greater than the total \gls{pv} generation:
\begin{align} 
p^\text{pv2bat}_{t,c} + p^\text{pv2l}_{t,c} + p^\text{pv2g}_{t,c} & \leq \sum_p gen^\text{pv}_{t,p,c} \label{constraint_pvbalance_gen}\\
p^\text{pv2g}_{t,c}, \; p^\text{pv2l}_{t,c} & \geq 0 \\
p^\text{pv2bat}_{t,c} & = p^\text{ch}_{t,c}
\end{align} 

Similarly, at each time step, the demand $l$ can be satisfied by: 1) power from \gls{pv} to the load $p^\text{pv2l}$; 2) power from the battery to the load $p^\text{bat2l}$ 
or 3) power from the grid to the load $p^\text{g2l}$. Mathematically,
\begin{align} 
p^\text{pv2l}_{t,c} + p^\text{bat2l}_{t,c} + p^\text{g2l}_{t,c} & \geq l_{t,c}  \label{constraint_pvbalance_load}\\
p^\text{bat2l}_{t,c}, \;p^\text{g2l}_{t,c} & \geq 0 \\
p^\text{bat2l}_{t,c} & = p^\text{dis}_{t,c}
\end{align} 

The self-consumed portion of the \gls{pv} generation $p^\text{sc}$ is defined as the total \gls{pv} electricity output that is directly or indirectly  consumed by the customer \cite{luthander2015photovoltaic}, which corresponds to the power from \gls{pv} to load and from battery to load, respectively, i.e.
\begin{align} 
& p^\text{sc}_{t,c} = p^\text{pv2l}_{t,c} + p^\text{bat2l}_{t,c}
\end{align}

\subsection{Formulation of optimization problem}
The objective is to optimize the investment and operational decisions of the \gls{pvb} system while minimizing the cost. 
The cost can be assessed using the discounted cash flow method, which calculates the \gls{npv} of the investment as the sum of investment costs and all discounted future cash flows. 

The total investment cost comprises the net \gls{pv} investment cost $C^\text{inv,pv}$ and the battery investment cost $C^\text{inv,bat}$. The \gls{pv} portion accounts for the investment subsidy $r^\text{sub,pv}$ and the tax rebate $r^\text{tax,pv}$ per kWp. The investment costs  across the five \gls{pv} categories is then given by 
\begin{align}\label{eq:InvCostPV}
& C^\text{inv,pv}_{c} = \sum_p (1-r^\text{tax,pv})(c^\text{inv,pv}_{p}-r^\text{sub,pv}_{p}) cap^\text{pv}_{p,c}
\end{align}
where $c^\text{inv,pv}$ is the cost of \gls{pv} per kWp for category $p$. The battery portion considers both the energy-related $c^\text{inv,bat-e}$ and the power-related $c^\text{inv,bat-p}$ investment costs, namely
\begin{align}\label{eq:InvCostBatt}
& C^\text{inv,bat}_{c} =  c^\text{inv,bat-e} cap^\text{bat-e}_{c}\!+\! c^\text{inv,bat-p} cap^\text{bat-p}_{c}
\end{align}

Future annual costs $C^\text{out}_{y,c}$ in year $y$ include both variable and fixed operational and maintenance costs of the \gls{pvb} system, i.e.,
\begin{align}
\begin{split}
& C^\text{out}_{y,c} = \sum_{t}(\sum_p c^\text{voc,pv}_p gen^\text{pv}_{t,p,c}+ c^\text{voc,bat-e} p^\text{dis}_{t,c}) \\
& \qquad \quad + c^\text{foc,bat-p} cap^\text{bat-p}_{c} 
\end{split}
\end{align}
where $c^\text{voc,pv}$, $c^\text{voc,bat-e}$ and $c^\text{foc,bat-p}$ are the variable cost parameter of \gls{pv}, along with the energy-related and the power-related cost parameters of the PV-battery.

The annual revenues $R^\text{in}$ include incomes from reimbursement of injecting electricity to the grid and savings from self consumption. To account for the degradation of the system, the annual revenues are multiplied by the annual system degradation rate $\delta^\text{deg}$ to the power of $y-y_0$, which is the difference between the considered year $y$ and the investment year of the \gls{pvb} system $y_0$. This results in the following equation:
\begin{align}
& R^\text{in}_{y,c} =  \sum_t (p^\text{p2g}_{t,c} pr^\text{inj}_{y,t,c} + p^\text{sc}_{t,c} pr^\text{retail}_{y,t,c})(1-\delta^\text{deg})^{y-y_0}
\end{align}
where $pr^\text{inj}$ and $pr^\text{retail}$ are the injection tariff and the retail electricity tariff. 
The savings from the self-consumed portion of the \gls{pv} generation in the model is calculated as the product of the self-consumed electricity and the retail electricity tariff, which better reflects the consumers’ savings and economic trade-offs. The retail electricity tariff is modeled using a dual tariff system with varying high and low tariffs depending on the corresponding annual electricity consumption category. Details of the retail electricity tariffs for the considered consumption categories L1-L11 are provided in Appendix~\ref{Apdx:App-elretailtariff}.

Furthermore, since the lifetime of the battery unit (i.e., 13 years) is shorter than that of the \gls{pvb} system (i.e., 30 years), a replacement of the battery unit and the possible residual value of the new battery unit at the end of the \gls{pvb} system needs to be accounted for. The replacement cost $C^\text{rpl,bat}_{y',c}$ in the year of replacement $y'$ is calculated using the investment cost in that year (i.e., $c^\text{inv,bat-e}_{y'}$ and $c^\text{inv,bat-p}_{y'}$), while the reinvested power and energy capacity of the battery is assumed to be the same as for the initial battery, i.e.
\begin{align}
C^\text{rpl,bat}_{y',c} & = c^\text{inv,bat-e}_{y'} cap^\text{bat-e}_{c}+c^\text{inv,bat-p}_{y'} cap^\text{bat-p}_{c} \\
\begin{split}
y' & = l^\text{bat}n'\!+\!1,  \\ 
n'& \in \{n'\!: n'\in \mathbb{Z}, 1 \leq n'\leq \lfloor (l^\text{sys}-1)/l^\text{bat}\rfloor \} 
\end{split}
\end{align}
where $l^\text{sys}$ and $l^\text{bat}$ are the lifetimes of the \gls{pvb} system and the battery. The number of needed battery replacements is calculated as $\lfloor (l^\text{sys}-1)/l^\text{bat}\rfloor$.

The residual value of the last reinvested battery $C^\text{res}$ is calculated as the multiplication of the annuity factor $\gamma^\text{ann}$, with the corresponding replacement cost and the residual battery lifetime by the end of the \gls{pvb} system calculated as $l^\text{bat-res}$: 
\begin{align}
    R^\text{res,bat}_{c} & = \gamma^\text{ann} l^\text{bat-res} C^\text{rpl,bat}_{y'=l^\text{bat}\lfloor (l^\text{sys}-1)/l^\text{bat}\rfloor+1,c} \\
    \gamma^\text{ann} & = \frac{wacc}{1-1/(1+wacc)^{l^\text{bat}}} \\
    l^\text{bat-res} & = [l^\text{bat}(\lfloor (l^\text{sys}-1)/l^\text{bat}\rfloor+1)  - l^\text{sys}]
    \label{eq_end}
\end{align}
where the year when the last required battery replacement takes place is $l^\text{bat}\lfloor (l^\text{sys}-1)/l^\text{bat}\rfloor+1$. For example, if battery lifetime (i.e., $l^\text{bat}$) is 13 years and the \gls{pvb} system lifetime (i.e., $l^\text{sys}$) is 30 years, then the number of needed battery replacements is calculated as two (i.e., $\lfloor (l^\text{sys}-1)/l^\text{bat}\rfloor$) and the year of the last required battery replacement is the 27th year (i.e., $l^\text{bat}\lfloor (l^\text{sys}-1)/l^\text{bat}\rfloor+1$) starting from the investment year.

Finally, the optimization problem for the entire lifetime of the \gls{pvb} system can be formulated as
\begin{equation}
\begin{aligned}
&\underset{}{\text{min}} ~~ \sum_{c} [C^\text{inv,pv}_{c} + C^\text{inv,bat}_{c} + \sum_{y=y0}^{l^\text{sys}} \frac{C^\text{out}_{y,c} -  R^\text{in}_{y,c}}{(1+wacc)^y} \nonumber \\
& \qquad \qquad +\sum_{n'=1}^{\lfloor (l^\text{sys}-1)/l^\text{bat}\rfloor }C^\text{rpl,bat}_{y'=l^\text{bat}n'+1,c} - R^\text{res,bat}_{c}] \nonumber\\ 	&\text{s.t.}
	\quad  \text{Constraints (\ref{eq_start})-(\ref{eq_end})}\nonumber
	\end{aligned}
\end{equation} 
where all future revenues and costs are discounted by \gls{wacc} to convert to the \gls{npv}.

\subsection{Technical and economic indicators}
Technical indicators for self-consumption rate and self-sufficiency rate as well as an economic indicator for payback period that will be used in the following analysis are described as follows:
\subsubsection{Self-consumption rate}
Based on definitions given in \cite{luthander2015photovoltaic}, the \gls{scr} is equal to the total \gls{pv} electricity output that is directly or indirectly consumed by the \gls{pvb} system owner divided by the total \gls{pv} generation.

\subsubsection{Self-sufficiency rate}
The \gls{ssr} represents the ratio of the electricity demand that can be satisfied by the \gls{pvb} system over the total electricity consumption of the \gls{pvb} system owner \cite{luthander2015photovoltaic}.

\subsubsection{Payback period}
The \gls{pbp} is defined as the investment cost divided by the yearly cash flow. The shorter the \gls{pbp} is, the more attractive the investment is.


\section{Results}
\label{sec:Results}
In the Baseline scenario, we run the model for each region and each customer group considering possible investments between 2020-2050 using a 5-year time step. More specifically, we run the static investment model 
for each investment year without considering any investments in previous years (i.e., a greenfield investment is simulated and the potentials for deployment are the same for each investment year).
Investment decisions are optimized by minimizing all investment and operating costs over a 30-year lifetime assumed for the \gls{pvb} system, where the operational decisions over all 8760 hours of the examined year are simulated and are assumed to be the same for the years of the remaining lifetime of the \gls{pvb} system. Different from the dynamic multi-period investment model that also optimizes investment timing and provides investment pathways, this work mainly aims to answer the question of how the economic viability of the \gls{pvb} system changes over time, i.e. for different investment years and its relation to the characteristics of different customer groups.
Each sensitivity scenario is only simulated for one example region (i.e., canton of Zurich) for the investment year 2050.

To better explain the results, in this section, we first show the results of an example customer group in Section~\ref{subsec:res_exHH}, then we illustrate the results for the example of the canton of Zurich in Section~\ref{subsec:res_exCanton}. Finally, the results at the national level (i.e., Switzerland) are analyzed in Section~\ref{subsec:res_CH}. For the first two subsections (i.e., Section~\ref{subsec:res_exHH} and Section~\ref{subsec:res_exCanton}), the Baseline results are presented first, followed by the results of the sensitivity analyses.

\subsection{Results for one representative customer group}
\label{subsec:res_exHH}
The average annual electricity consumption per household in Switzerland is 5000 kWh \cite{bundesamt2019schweizerische} and the average annual solar irradiance in Switzerland is 1267 kWh/m$^2$ \cite{Bauer2019}.
To represent an average customer group in Switzerland, we select the group with the following criteria: canton of Zurich (REG1), rooftop size of 108-120 m$^2$ (A13), annual irradiation of 1150-1300 kWh/m$^2$/year (IRR2) and electricity consumption of 4500-5500 kWh/year (L5).
As mentioned in Section~\ref{subsec:Data_B}, each customer group is represented using the median values of the rooftop size, the annual irradiation and the electricity consumption from within the group. 
Since a range of rooftop sizes in a particular customer group are analyzed together using representative characteristics, the investment decision for each group yields a single combination of \gls{pv} and battery investments for all rooftops within this group.
For example, the selected customer group has a median annual electricity consumption of 5025 kWh, a median annual solar irradiation of 1212 kWh/m$^2$ and a median rooftop size of 113 m$^2$ (i.e., equivalent to 18.8 kWp potential of \gls{pv}). 
The aggregated rooftop area within the considered customer group is equal to 
20'751 m$^2$, which means the optimized decision for the representative customer 
is reflective of around 184 customers (i.e., total rooftop size divided by the median rooftop size of the customer group).
Note that the results shown in this section are only for the single representative rooftop within the single selected customer group.

\subsubsection{Baseline results - investment}
Table~\ref{tab:results_base_Zurich_exHH_base} shows the optimal investment decisions of the example customer group over the simulation horizon (i.e. 2020-2050) for the Baseline scenario.
Comparing the results over the years, the optimal \gls{pv} and battery sizes for the representative rooftop in this customer group continue to increase. The \gls{pbp} in general follows a decreasing trend from above 13 years in 2025 to below 10 years in 2050 except for an increase from 2030 to 2035, which is mainly due to the subsidy expiration by the end of 2030. Correspondingly, the \gls{npv} in general increases over time except a slight decrease from 2030 to 2035.
Changes to these optimal investment decisions and the resulting \gls{pbp} and \gls{npv} over the years can be mainly traced back to the decreasing \gls{pv} and battery costs and the increasing retail electricity tariffs. The \gls{pvb} C-rate is fairly consistent over the years between 0.19-0.23, which is reasonable considering the popular household consumer solar battery systems available nowadays (e.g. the 13.5 kWh/3.6 kW Tesla Powerwall2 with a C-rate of 0.27 \cite{web_tesla} and the 15 kWh/3.3 kW Sonnenbatterie Eco9 with a C-rate of 0.22 \cite{web_sonnenbatterie}).
Furthermore, the \gls{ssr} increases with the increasing size of the \gls{pvb} system, meaning that the homeowner is able to supply more and more of its own demand.
In contrast, the \gls{scr} first increases and then decreases, indicating that the larger \gls{pvb} systems tend to sell a larger portion of their production to the grid. This result also shows that the investment profitability in early years is driven by the high \gls{scr} while further into the future it is instead driven by the decreasing costs. In these future years, it is also profitable to install a \gls{pvb} system that is larger than required for the consumers' demand.  
\begin{table}[t]
\renewcommand{\arraystretch}{1.1}
\centering
\caption{Baseline analysis for the representative rooftop of the example customer group.}
\label{tab:results_base_Zurich_exHH_base}
\scriptsize
\begin{tabular}{|l|c|c|c|c|c|c|c|}
\hline
\multicolumn{1}{|c|}{\multirow{2}{*}{\textbf{Year}}} &  \multicolumn{3}{c|}{\textbf{\begin{tabular}[c]{@{}c@{}}Investment size\\ {[}kW or kWh{]}\end{tabular}}} & \multirow{2}{*}{\textbf{\begin{tabular}[c]{@{}c@{}}NPV\\
{[}kEUR{]}\end{tabular}}} &
\multirow{2}{*}{\textbf{\begin{tabular}[c]{@{}c@{}}PBP\\ {[}Year{]}\end{tabular}}} & \multirow{2}{*}{\textbf{SCR}} & \multirow{2}{*}{\textbf{SSR}}\\ \cline{2-4}
\multicolumn{1}{|c|}{} & \textbf{PV} & \begin{tabular}[c]{@{}c@{}}\textbf{BESS-e} \\/ \textbf{BESS-p} \end{tabular}& \textbf{\begin{tabular}[c]{@{}c@{}}BESS\\ C-rate\end{tabular}} & & & & \\ \hline
\textbf{2020} & 0 & 0 / 0 & n/a & n/a & n/a & n/a & n/a\\ \hline
\textbf{2025} & 2.0 & 3.0 / 0.6 & 0.20 & 0.5 & 13.5 & 74\% & 29\%\\ \hline
\textbf{2030} & 2.3 & 5.7 / 1.0 & 0.19 & 1.4 & 11.8 & 80\% & 36\%\\ \hline
\textbf{2035} & 2.7 & 7.2 / 1.4 & 0.20 & 1.3 & 13.0 & 80\% & 42\%\\ \hline
\textbf{2040} & 3.3 & 8.6 / 1.8 & 0.21 & 2.3 & 11.7 & 77\% & 48\%\\ \hline
\textbf{2045} & 6.0 & 10.0 / 2.3 & 0.23 & 3.4 & 11.4 & 56\% & 64\%\\ \hline
\textbf{2050} & 6.0 & 10.3 / 2.3 & 0.22 & 5.1 & 9.5 & 56\% & 65\%\\ \hline
\end{tabular}
\vspace{-0.2cm}
\end{table}

\subsubsection{Baseline results - dispatch}
\begin{figure}[htbp]
\vspace{-0.2cm}
    \centering
    \begin{subfigure}[b]{0.535\textwidth} 
      \includegraphics[width=\textwidth]{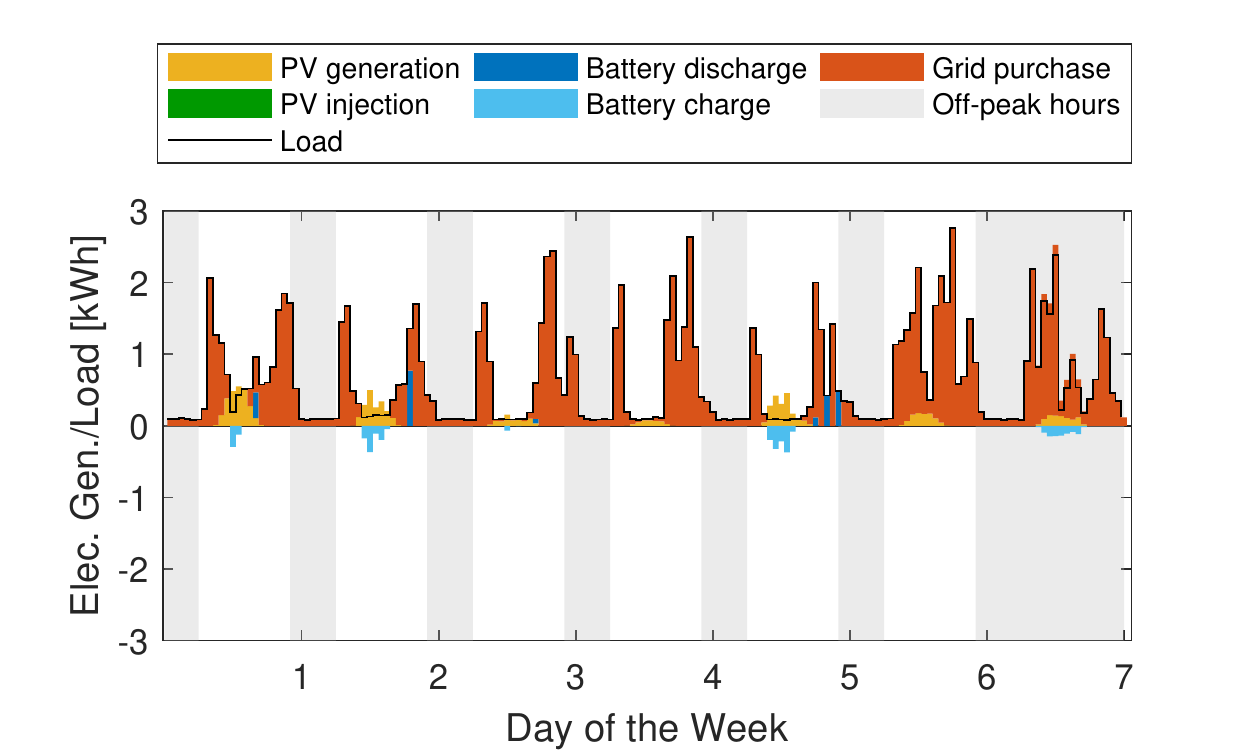}
      \caption{Winter - 2030}
      \label{fig_ExHHdispatch_2030_winter}
    \end{subfigure}    \begin{subfigure}[b]{0.535\textwidth} 
      \includegraphics[width=\textwidth]{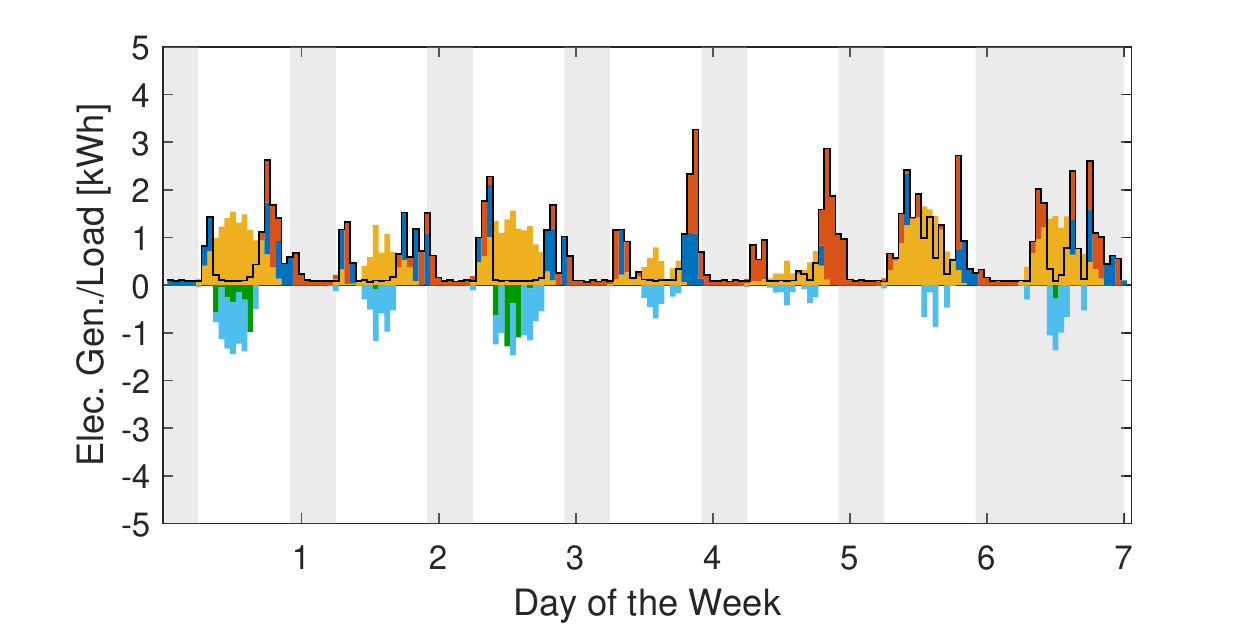}
      \caption{Summer - 2030}
      \label{fig_ExHHdispatch_2030_summer}
    \end{subfigure}
    \begin{subfigure}[b]{0.535\textwidth} 
      \includegraphics[width=\textwidth]{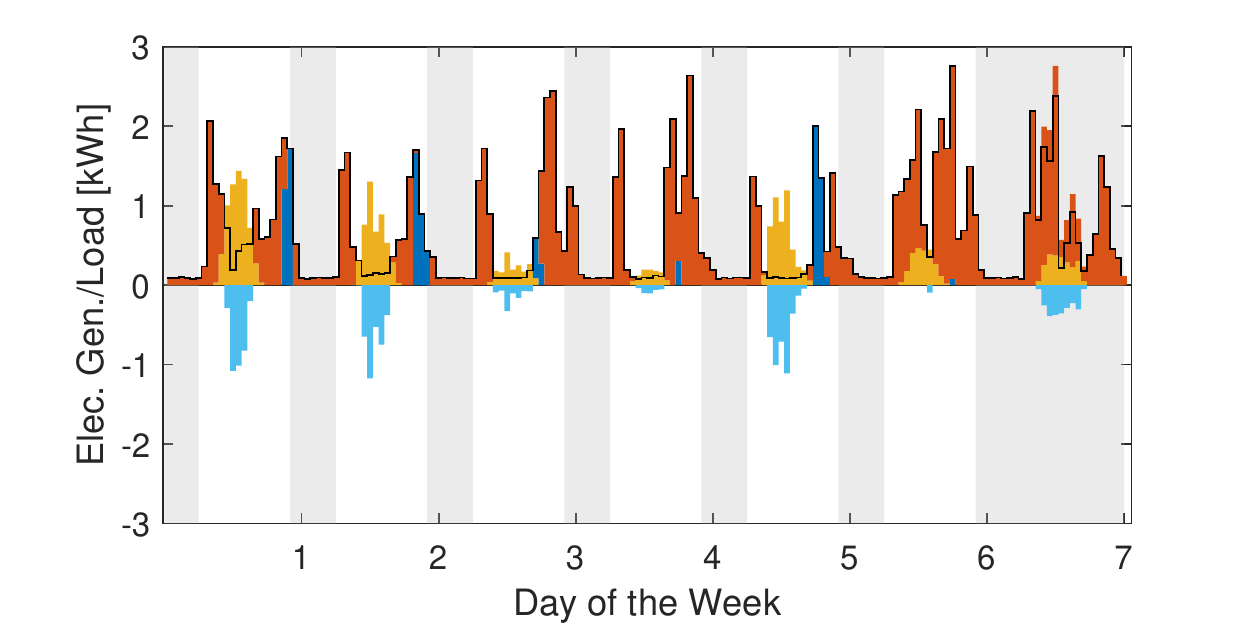}
      \caption{Winter - 2050}
      \label{fig_ExHHdispatch_2050_winter}
    \end{subfigure}
    \begin{subfigure}[b]{0.535\textwidth} 
      \includegraphics[width=\textwidth]{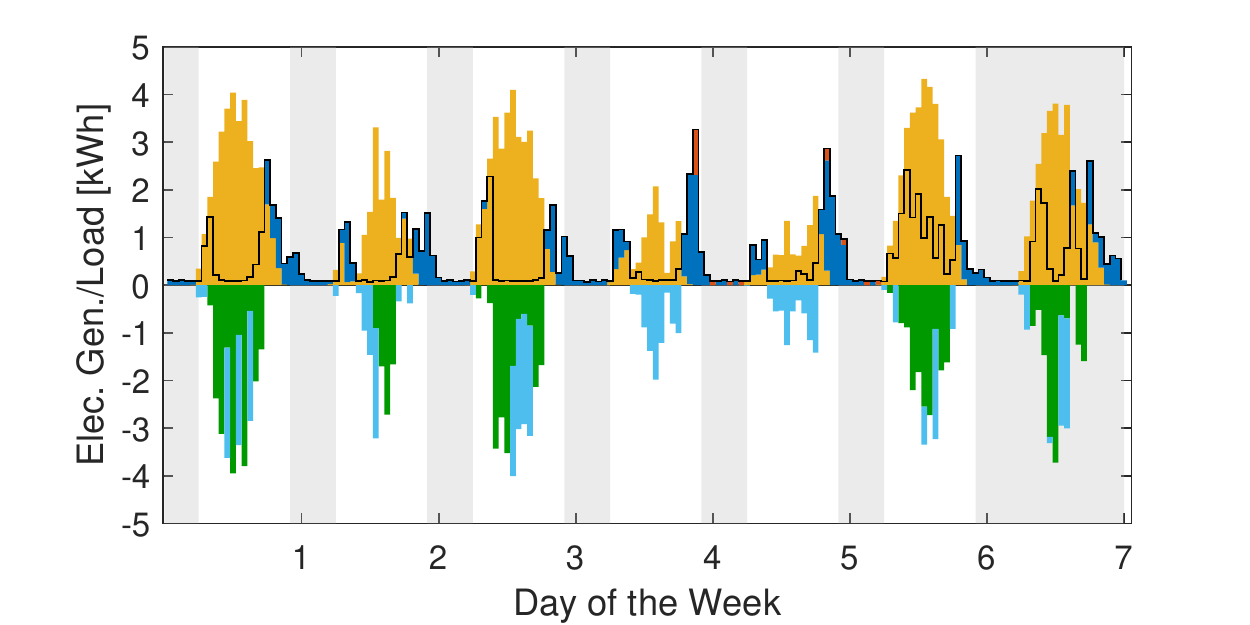}
      \caption{Summer - 2050}
      \label{fig_ExHHdispatch_2050_summer}
    \end{subfigure}
      \caption{Dispatch of the PVB system for the representative rooftop of the example customer group in 2030 and 2050.}\label{fig:dispatch_exHH}
\end{figure}

Figure~\ref{fig:dispatch_exHH} shows the generation and load dispatch of the \gls{pvb} system of an example winter and summer week for 2030 and 2050, respectively. Both the selected winter and summer weeks start from a Monday. Low electricity tariff hours (i.e., off-peak hours) are marked by the gray area, while the rest is the high electricity tariff period.

In general, the battery discharges/charges when the load is higher/lower than the \gls{pv} generation to increase the self-consumption rate and in turn improve the profitability of the \gls{pvb} system investment. An exception can be observed on the 7th day (i.e., Sunday) of the winter weeks, when the battery charges even though the load is higher than the \gls{pv} generation. This is due to the assumption that all hours on Sunday are low electricity tariff hours (i.e., off-peak). The \gls{pvb} system therefore takes advantage of the cheap electricity from the grid to supply the demand while the \gls{pv}-battery absorbs the \gls{pv} generation for later use during high electricity tariff hours. 
Furthermore, discharging is ideally done during the peak electricity tariff hours (i.e., 6:00-22:00 from Monday to Saturday) in order to reduce the electricity bill. 
Note that in the Baseline scenario, the retail electricity tariffs are modeled using a dual system while the injection tariff is assumed to be constant over all hours. Hence, non-unique solutions might occur as the charging/discharging in different hours in the same price tier could result in the same objective value. However, this is irrelevant for our study.

As shown in Table~\ref{tab:results_base_Zurich_exHH_base}, the optimal invested battery size increases from 5.7 kWh/1.0 kW in 2030 to 10.3 kWh/2.3 kW in 2050, whereas the optimal \gls{pv} size increases from 2.3 kW to 6.0 kW between the same years. 
Comparing the 2050 dispatch results to that of 2030 both shown in Fig.~\ref{fig:dispatch_exHH}, the grid purchases decrease while the \gls{pv} injections increase due to the larger size of the installed \gls{pv} system. Although the battery size is also expanded, the general pattern of the \gls{pvb} system behavior does not change significantly. Additionally, the dynamics of the power consumed/sold to the grid are exacerbated since in 2050 the installed battery capacity per kW of \gls{pv} is lower.

\subsubsection{Sensitivity scenario results}
\begin{table}[t]
\renewcommand{\arraystretch}{1.1}
\centering
\caption{Sensitivity analysis for the representative rooftop of the example customer group in 2050.}
\label{tab:results_base_Zurich_exHH_sensitivity}
\scriptsize
\begin{tabular}{|l|c|c|c|c|c|c|c|}
\hline
\multicolumn{1}{|c|}{\multirow{2}{*}{\textbf{Year}}} &  \multicolumn{3}{c|}{\textbf{\begin{tabular}[c]{@{}c@{}}Investment size\\ {[}kW or kWh{]}\end{tabular}}} & \multirow{2}{*}{\textbf{\begin{tabular}[c]{@{}c@{}}NPV\\
{[}kEUR{]}\end{tabular}}} &
\multirow{2}{*}{\textbf{\begin{tabular}[c]{@{}c@{}} PBP\\ {[}Year{]}\end{tabular}}} & 
\multirow{2}{*}{\textbf{SCR}} &
\multirow{2}{*}{\textbf{SSR}} \\ \cline{2-4}
\multicolumn{1}{|c|}{} & \textbf{PV} & \begin{tabular}[c]{@{}c@{}}\textbf{BESS-e} \\/ \textbf{BESS-p} \end{tabular} & \textbf{\begin{tabular}[c]{@{}c@{}}BESS\\ C-rate\end{tabular}} & & & & \\ \hline
\textbf{Base.} & 6.0 & 10.3 / 2.3 & 0.22 & 5.1 & 9.5 & 56\% & 65\%\\ \hline
\textbf{SC1} & 6.0 & 8.6 / 1.9 & 0.23 & 3.2 & 11.8 & 55\% & 63\%\\ \hline
\textbf{SC2} & 6.0 & 14.4 / 2.7 & 0.19 & 7.2 & 7.0 & 57\% & 67\%\\ \hline
\textbf{SL} & 6.0 & 7.1 / 1.4 & 0.20 & 5.4 & 8.9 & 57\% & 65\%\\ \hline
\textbf{SP1} & 2.1 & 5.6 / 1.1 & 0.19 & 0.6 & 14.7 & 82\% & 34\%\\ \hline
\textbf{SP2} & 6.0 & 8.6 / 2.1 & 0.24 & 1.0 & 14.8 & 54\% & 63\%\\ \hline
\textbf{SP3} & 6.0 & 7.9 / 2.0 & 0.25 & 2.0 & 12.9 & 54\% & 62\%\\ \hline
\textbf{SP4} & 6.0 & 10.5 / 2.4 & 0.22 & 4.7 & 9.9 & 56\% & 65\%\\ \hline
\textbf{SP5} & 6.0 & 10.3 / 2.4 & 0.23 & 5.4 & 9.3 & 56\% & 65\%\\ \hline
\textbf{SP6} & 10.0 & 10.6 / 2.6 & 0.24 & 7.0 & 9.6 & 39\% & 75\%\\ \hline
\textbf{SP7} & 6.0 & 11.8 / 2.6 & 0.22 & 11.6 & 6.2 & 57\% & 66\%\\ \hline
\textbf{SP8} & 10.0 & 12.5 / 2.8 & 0.22 & 12.6 & 7.2 & 39\% & 76\%\\ \hline
\textbf{SP9} & 12.3 & 12.2 / 3.0 & 0.24 & 15.1 & 7.2 & 34\% & 79\%\\  \hline
\textbf{SP10} & 6.0 & 10.6 / 2.4 & 0.22 & 4.3 & 10.2 &56\% & 65\%\\ \hline
\textbf{SW1} & 6.0 & 10.5 / 2.3 & 0.22 & 8.3 & 9.8 & 56\% & 65\%\\ \hline
\textbf{SW2} & 6.0 & 10.5 / 2.3 & 0.22 & 1.6 & 8.7 & 56\% & 65\%\\ \hline
\textbf{SB1} & 6.0 & 7.6 / 1.8 & 0.23 & 3.7 & 11.0 & 53\% & 62\%\\ \hline
\textbf{SB2} & 2.0 &  n/a & n/a & 0.9 & 11.9 & 49\% & 19\%\\ \hline
\end{tabular}
\vspace{-0.2cm}
\end{table}
The results of simulating different sensitivity scenarios in 2050 are provided in Table~\ref{tab:results_base_Zurich_exHH_sensitivity}. The main observations are:
\begin{itemize}
    \item \underline{Cost sensitivity}: The optimal battery size and the \gls{npv} decreases/increases,  and the \gls{pbp} increases/decreases in the high/low cost scenario (i.e., SC1/SC2), while the optimal \gls{pv} size is unchanged. This is due to the fact that the future battery cost is subject to higher uncertainties than that of \gls{pv}.
    \item \underline{Load sensitivity}: When applying the aggregate load profile (i.e. SL) with equal energy consumed, the optimal \gls{pv} size stays unchanged,
    but both the optimal battery size and battery C-rate are reduced. This is because the aggregate load profile is flatter and better matches the \gls{pv} generation profile than the individual load profiles, therefore a smaller battery is required to achieve similar \gls{scr} and \gls{ssr} values to those of the Baseline scenario, which results in a higher \gls{npv} and a shorter \gls{pbp}.
    \item \underline{Price sensitivity I}: 
    Having access to the hourly wholesale market (i.e. SP1-SP9) has mixed impacts on the investment decisions, the \gls{npv} and the \gls{pbp}, depending on how the retail and wholesale electricity prices evolve.
    \subitem Comparing the results under the same retail electricity price (i.e., SP1 vs. SP2 vs. SP3; SP4 vs. SP5 vs. SP6; SP7 vs. SP8 vs. SP9), higher wholesale market prices increase the optimal \gls{pv} investment size and the \gls{npv}, and reduce the \gls{scr} since it means greater revenues for the same amount of electricity injection.
    However, higher wholesale prices in general reduce the optimal battery (energy and power) capacity invested per unit installed \gls{pv} capacity. This is because the spread between wholesale and retail electricity prices is smaller when higher wholesale electricity is simulated, which lowers the savings earned by using batteries. 
    Interestingly, the battery C-rate increases with the increasing wholesale price development (i.e., from SP1 to SP3, from SP4 to SP6 and from SP7 to SP9) since higher wholesale prices encourage investments in a larger \gls{pv} unit, which in turn requires a higher C-rate to cope with the increased dynamics of the net load. 
    \subitem Comparing the results under the same wholesale electricity price (i.e., SP1 vs. SP4 vs. SP7; SP2 vs. SP5 vs. SP8; SP3 vs. SP6 vs. SP9), the higher retail electricity prices (i.e., SP7-SP9) reduce the \gls{pbp} and increase the \gls{npv} and the optimal size of both \gls{pv} and battery units.
    \subitem  The impact of the wholesale electricity price is limited compared to the influence of the retail electricity tariff as in general the retail electricity price level is higher than the wholesale electricity price.
    \item \underline{Price sensitivity II}: When the injection tariff is zero and no wholesale market access is granted (i.e. SP10), the optimal \gls{pv} size is the same but the battery size is slightly higher. The resulting \gls{npv} decreases and the \gls{pbp} increases slightly compared to the Baseline scenario, which shows the limited impact of injection tariffs in 2050 for the example of the considered customer group.
    \item \underline{WACC sensitivity}: Increasing the value of the \gls{wacc} from 4\% (i.e., Baseline) to 8\% (i.e., SW1) or reducing it to 2\% (i.e., SW2) does not impact the invested \gls{pv} and battery sizes and only slightly changes the \gls{pbp}. However, the \gls{npv} varies significantly under different assumptions of \gls{wacc} because of the discounting factor of future cash flows.
    \item \underline{Battery price sensitivity}: Adjusting the battery price using the current Tesla Powerwall 2 cost in Switzerland (i.e., SB1) results in even less battery investments than the high cost scenario SC1, which highlights the importance of considering regional differences of the \gls{pvb} system investment costs.
    \item \underline{Battery integration sensitivity}: When no battery installation is considered (i.e., SB2), the \gls{npv} is much lower and the \gls{pbp} is longer than in the Baseline scenario, which shows that the successful combination of battery units with \gls{pv} does contribute to increasing the profitability of the \gls{pvb} system for the example customer group in 2050. 
\end{itemize}
The \gls{npv} and the \gls{pbp} are subject to future uncertainties and vary greatly between different sensitivity simulation scenarios. The economic viability of the \gls{pvb} system is especially sensitive to the future cost of \gls{pv} and battery and the electricity price development. 
  
\subsection{Results for all customer groups within the canton of Zurich}
\label{subsec:res_exCanton}
To broaden the scope of the results, this subsection discusses the resulting optimal investment decisions for all 2200 customer groups in the canton of Zurich. The combination of these customer groups represents 435'815 individual consumers/households and a combined rooftop space of 28.4 km$^2$, which is equivalent to a cumulative \gls{pv} potential of 4.7 GW.

\subsubsection{Baseline results}
\begin{table}[b]
\renewcommand{\arraystretch}{1.1}
\centering
\caption{Baseline result analysis for canton Zurich, years 2020-2050.}
\label{tab:results_base_Zurich}
\scriptsize
\begin{tabular}{|l|c|c|c|c|c|c|c|}
\hline
\multirow{3}{*}{\textbf{Year}} &  \multicolumn{3}{c|}{\textbf{\begin{tabular}[c]{@{}c@{}}WAVG \\ investment size\\ {[}kW or kWh{]}\end{tabular}}} & \multirow{3}{*}{\textbf{\begin{tabular}[c]{@{}c@{}}WAVG \\ NPV \\ {[}kEUR{]}\end{tabular}}} & \multirow{3}{*}{\textbf{\begin{tabular}[c]{@{}c@{}}WAVG \\ PBP\\ {[}Year{]}\end{tabular}}} & \multirow{3}{*}{\textbf{\begin{tabular}[c]{@{}c@{}}Cum.\\PV \\ {[}GW{]}\end{tabular}}}
& \multirow{3}{*}{\textbf{\begin{tabular}[c]{@{}c@{}}Cum.\\BESS \\ {[}GWh\\/ GW{]}\end{tabular}}}
\\ \cline{2-4}
\multicolumn{1}{|c|}{} & \textbf{PV} & \begin{tabular}[c]{@{}c@{}} \textbf{BESS-e} \\ / \textbf{BESS-p} \end{tabular} & \textbf{\begin{tabular}[c]{@{}c@{}}BESS\\ C-rate\end{tabular}} & & & & \\ \hline
\textbf{2020} & 4.5 & 1.1 / 0.3 & 0.25 & 2.3 & 11.5 & 1.4 & 0.4 / 0.1 \\ \hline
\textbf{2025} & 5.8 & 7.0 / 1.5 & 0.21 & 3.3 & 11.7 & 2.2 & 2.7 / 0.6 \\ \hline
\textbf{2030} & 7.6 & 14.0 / 2.8 & 0.20 & 6.0 & 11.0 & 3.0& 5.4 / 1.1\\ \hline
\textbf{2035} & 7.8 & 16.7 / 3.4 & 0.20 & 6.5 & 11.6 & 3.0 & 6.4 / 1.3 \\ \hline
\textbf{2040} & 8.3 & 18.1 / 3.7 & 0.20 & 8.9  & 10.2 & 3.2 & 7.0 / 1.4  \\ \hline
\textbf{2045} & 8.8 & 18.9 / 3.9 & 0.20 & 10.9 & 9.3 & 3.4 & 7.3 / 1.5 \\ \hline
\textbf{2050} & 9.2 & 19.8 / 4.0 & 0.20 & 13.3 & 8.3 & 3.6 & 7.7 / 1.6  \\ \hline
\end{tabular}
\vspace{-0.2cm}
\end{table}
Table~\ref{tab:results_base_Zurich} shows the weighted average size, \gls{npv} and \gls{pbp} as well as the cumulative capacity of the \gls{pvb} investments across all customer groups in the canton of Zurich. The assigned weights are the number of customers (i.e., rooftops) in each customer group.
Different from the results of the example customer group, it is profitable to invest in \gls{pv} and PV-battery for some customer groups already in the current year (i.e., 2020) in Zurich.
Moving from 2020 to 2050, the weighted average size of the invested \gls{pv} and battery units is increasing, mainly as a result of the decreasing costs. This result is consistent with the observation drawn from the previous results of the example customer group.
This growth is prominent for the battery during the period between 2020 and 2035, when the estimated battery price drops significantly (for more details see Appendix~\ref{Apdx:App-futurecosts_battery}).
Although the \gls{npv} increases over the years, the weighted average \gls{pbp} fluctuates between 2020 and 2035 and decreases afterwards, which is due to the mixed impacts of the investment subsidy decrease, the injection tariff variation, the retail tariff increase and the investment cost decrease. In other words, the annual net cash inflow does not increase as much as the investment cost during this period (i.e., 2020-2035). 
Individual impacts of some of these important input factors will be further investigated later using sensitivity analysis.
Similar to the trend of the weighted average investment capacity, the cumulative \gls{pv} and battery investment capacities also increase over time from 1.4 GW and 0.4 GWh/0.1 GW in 2020 to 3.6 GW and 7.7 GWh/1.6 GW in 2050, while the total \gls{pv} deployment potential modeled for the canton of Zurich is 4.7 GW.
It is worth noting that the resulting investment capacities account for all investments that could achieve positive \gls{npv}, even if small, over the 30-year lifetime of the \gls{pvb} system, while in reality investors might have higher expectations for the \gls{npv} and the \gls{pbp}.
\begin{figure}[t]
    \centering
    \includegraphics[width=0.5\textwidth]{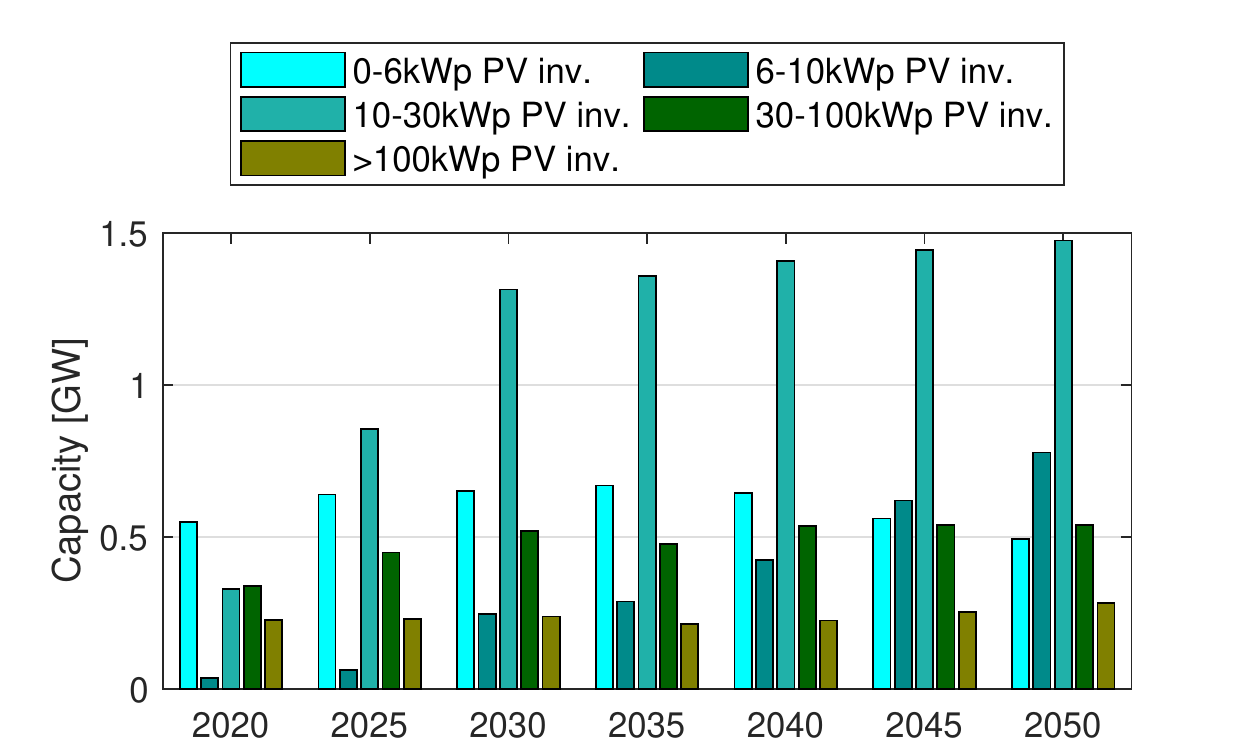}
    \caption{Cumulative PV investments in different size categories for the canton Zurich, years 2020-2050.}\label{fig:inv_5pv_exCanton}
\end{figure}

Figure~\ref{fig:inv_5pv_exCanton} shows the cumulative \gls{pv} investments in different size categories from 2020 to 2050 for the canton of Zurich. Please note by cumulative, we refer to the summation over all customer groups in any particular year and not over time as we start with a greenfield in every considered year. While the cumulative investments in 6-10 kW and 10-30 kW \gls{pv} units increase significantly from 2020 to 2050, investments in other \gls{pv} size categories fluctuate over the years: a) investments in \gls{pv} sizes below 6 kW first increase then decrease, which is likely due to the fact that investments are driven by high \gls{scr} in early years and most customer groups install smaller \gls{pv} units that do not fully exploit the potential of their rooftop sizes\footnote{Assuming the considered available rooftop potential in the canton of Zurich is fully exploited 
(i.e., \gls{pv} units sizes are maximized for every single house based solely on the corresponding rooftop size),
then the maximum investments in 0-6 kW, 6-10 kW, 10-30 kW, 30-100 kW and >100 kW \gls{pv} units are 0.53 GW, 0.77 GW, 2.29 GW, 0.82 GW and 0.33 GW, respectively.}. Since the optimal \gls{pv} size increases mainly as a result of the decreasing investment cost, the cumulative \gls{pv} investment capacity gradually shifts from smaller to greater \gls{pv} size categories; b) investments in \gls{pv} above 30 kW in general increase over time except for a small decrease between 2030 and 2035, which shows that compared to that of smaller \gls{pv} units the economic viability of larger \gls{pv} units relies more on the investment subsidy. 

\begin{figure}[t]
    \centering
    \begin{subfigure}[b]{0.5\textwidth} 
      \includegraphics[width=\textwidth]{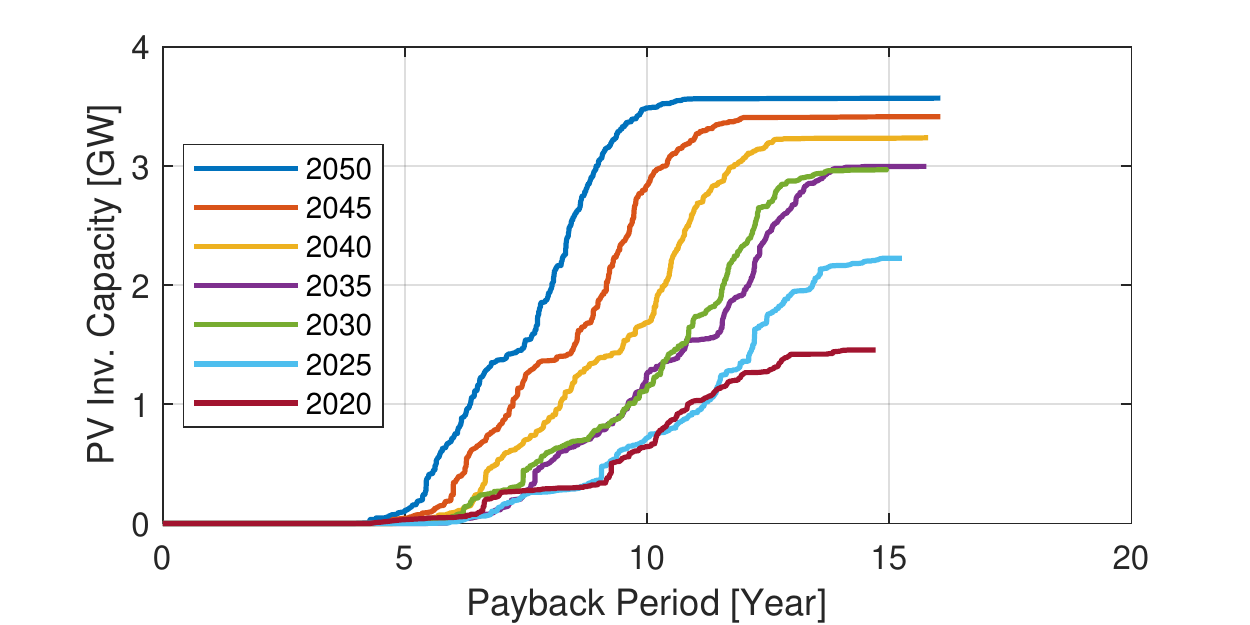}
      \caption{PV investment capacity}
      \label{fig_Baseline_PBPdistribution_ZH_pv}
    \end{subfigure}    
    \begin{subfigure}[b]{0.5\textwidth} 
      \includegraphics[width=\textwidth]{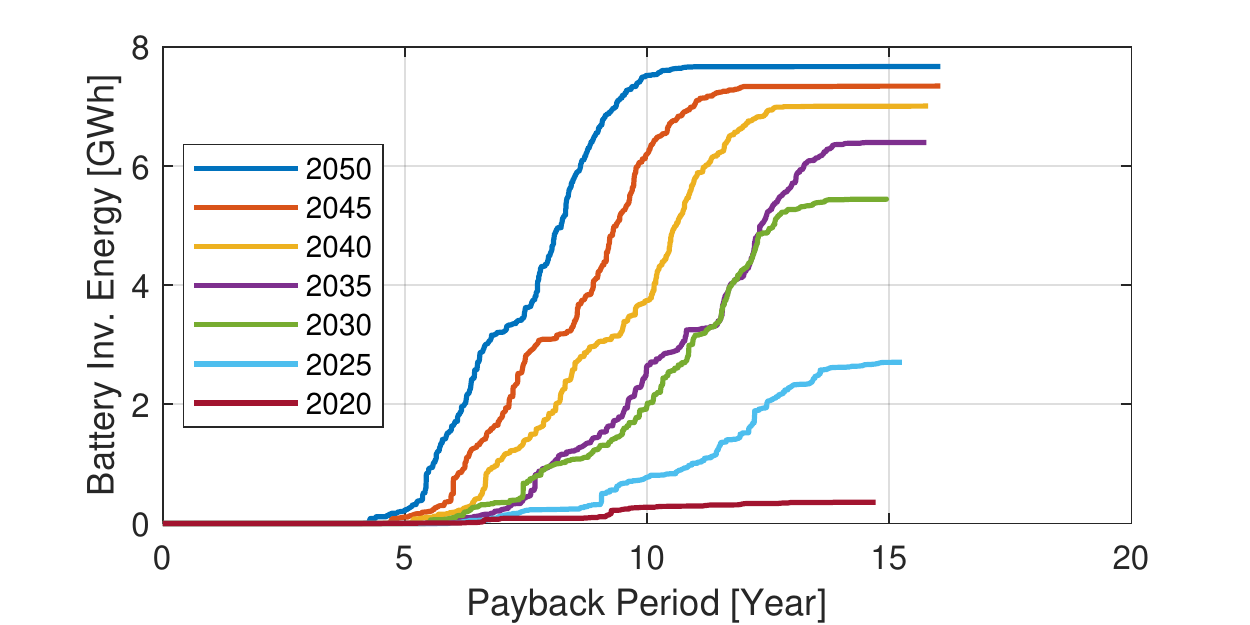}
      \caption{Battery investment energy capacity}
      \label{fig_Baseline_PBPdistribution_BESSe_ZH}
    \end{subfigure}  
    \begin{subfigure}[b]{0.5\textwidth} 
      \includegraphics[width=\textwidth]{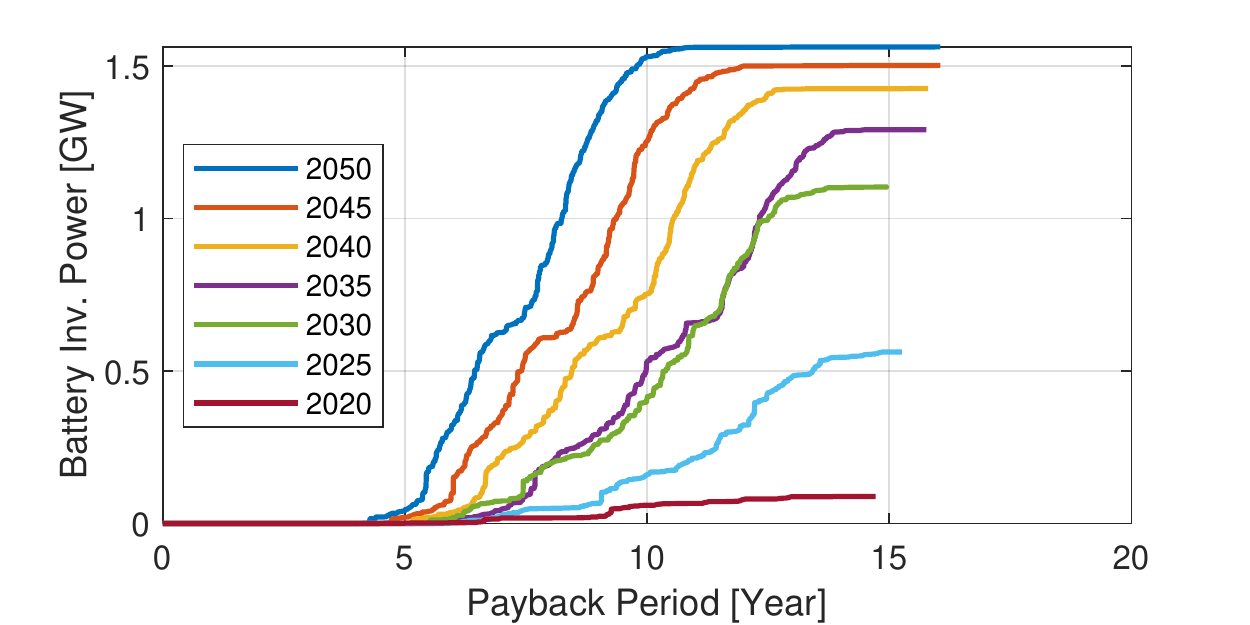}
      \caption{Battery investment power capacity}
      \label{fig_Baseline_PBPdistribution_BESSp_ZH}
    \end{subfigure} 
    \caption{Optimal investment against \gls{pbp} of the Baseline scenario of 2020-2050 for the canton of Zurich.}\label{fig:Baseline_PBPdistribution_ZH}
\vspace{-0.3cm}
\end{figure}

Figure~\ref{fig:Baseline_PBPdistribution_ZH} depicts the relationship between the total optimal \gls{pv} and battery investment capacities and the \gls{pbp} over all 2200 customer groups. In all three plots, each line represents the accumulated capacity of the 2200 customer groups, which have been ordered by increasing \glspl{pbp}. In general, the total capacities of the invested \gls{pv} and battery units increase over the years, along with yielding more capacities that have shorter \gls{pbp}. However, the curves of 2030 and 2035 (especially for the \gls{pv} investment capacity) intersect/overlap, which is, as elaborated already also earlier, mainly because of the mixed effects of cost reductions and the investment subsidy expiration by the end of 2030. 

\subsubsection{Sensitivity scenario results}
\begin{figure}[t]
  \centering
    \includegraphics[width=.45\textwidth]{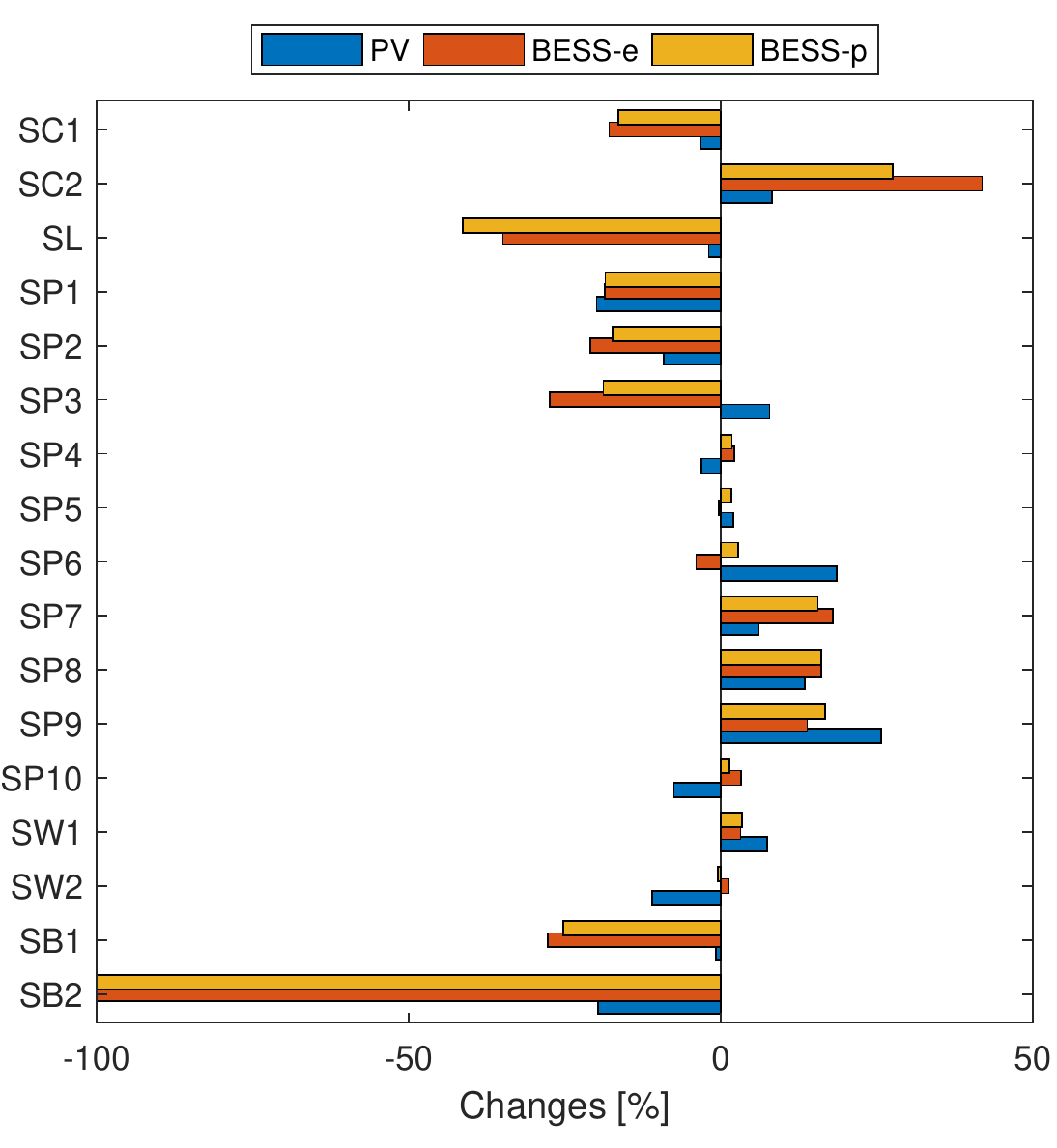}
    \caption{Investment changes in the example of the canton of Zurich under different scenarios.}
    \label{fig:sensitivity_ZH}
\vspace{-0.3cm}
\end{figure}

Figure~\ref{fig:sensitivity_ZH} illustrates the changes compared to the Baseline of the total investment capacities of \gls{pv} and batteries in the canton of Zurich for each sensitivity scenario in 2050. 
Focusing on the \gls{pv} investment results, several different scenarios result in a similar cumulative \gls{pv} capacity to the Baseline, (i.e. costs SC1-SC2, load profiles SL and \gls{wacc} values SW1-SW2).
In contrast, the optimal \gls{pv} investment capacity is highly sensitive to the electricity price developments (i.e., SP1-SP10), with the lowest/highest price scenario (i.e., SP1/SP9) yielding the lowest/highest level of \gls{pv} integration. 
Alternatively, the cumulative battery energy and power capacities vary significantly among scenarios, with the lowest battery capacity invested in the aggregated load scenario SL and the highest battery capacity invested in the low cost scenario SC2.
Considering price scenarios SP1-SP10, the highest total battery investment capacity is obtained under the price scenario SP7 (i.e., highest retail price increase of 2\%/year and lowest wholesale price increase of -1\%/year),  
while the lowest battery investment is obtained under the price scenario SP3 (i.e., lowest retail price increase of 0\%/year and highest wholesale price increase of 3\%/year). This is likely due to the fact that a smaller spread between wholesale and retail electricity prices in turn decreases the profitability of battery investments in 2050, when in general more \gls{pv} is invested than required for the consumers' demand and the battery investment is driven more by shifting the \gls{pv} injection from low to high wholesale electricity price hours than to increase the \gls{scr}.  

\subsection{Results for Switzerland}
\label{subsec:res_CH}
In this section, we only show results for the Baseline scenario and for years 2020, 2030, 2040 and 2050. The results consider all 26 regions (i.e., cantons) in Switzerland with 2200 customer groups within each region. The combination of all these customer groups represents 3’795’145 individual consumers/households and a rooftop area of 224 km$^2$, which is equivalent to a cumulative \gls{pv} potential of 37 GW.

Since the investment decisions are optimized by maximizing the \gls{npv} of the investment, the resulting \gls{pbp} could be up to the lifetime of the \gls{pvb} system (i.e., 30 years). However, most investors would expect a \gls{pbp} that is much shorter than the lifetime of the \gls{pvb} system.
The \gls{pbp} of the currently installed \gls{pvb} systems varies across countries, locations and customer groups.
A recent study \cite{web_solarchoice} conducted in Australia shows a \gls{pbp} of 5 to 12 years, whereas some research \cite{web_energyhub} suggests that the \gls{pbp} could be as long as 16 years.

Investments that result in long \glspl{pbp} are likely not of high interest to customers. We therefore focus on two cases and define them as follows:
\begin{itemize}
    \item \textbf{Fast recoverable} investment: \gls{pbp} is less than 10 years;
    \item \textbf{Moderately fast recoverable} investment: \gls{pbp} is less than 15 years.
\end{itemize}

\subsubsection{Baseline results - investment}
\begin{figure}[t]
\vspace{-0.2cm}
  \centering
  \includegraphics[width=0.5\textwidth]{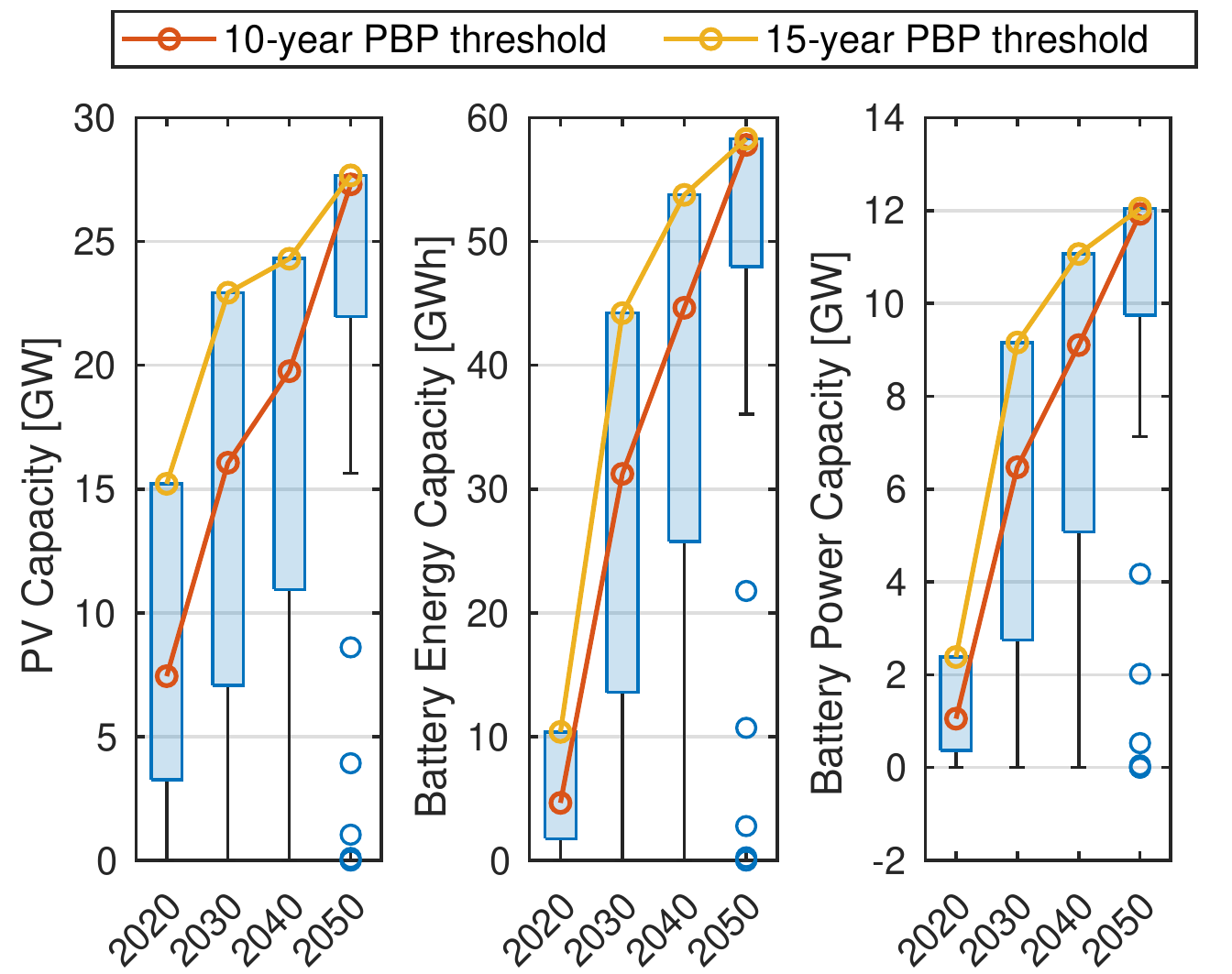}
  \label{fig_Baseline_CH_yearly_pv}
\vspace{-0.3cm}
  \caption{Optimal yearly investment under different \glspl{pbp}.}\label{fig:Baseline_yearly_CH}
\vspace{-0.2cm}
\end{figure}
Figure~\ref{fig:Baseline_yearly_CH} shows the optimal investments in Switzerland between 2020-2050. 
Each year is represented by a whisker plot where each value within this plot represents the cumulative capacity that is built in Switzerland with a \gls{pbp} from zero to 30 years. 
The investment decision is highly sensitive to the acceptable \gls{pbp}, especially between 2030 and 2040, when the cost reduction is not high enough to achieve a short \gls{pbp} for all customer groups. 
It can be observed that for each year, the resulting 15-year \gls{pbp} investment results are almost always on the top edge of the box. This can be explained by the fact that most of the investments have a \gls{pbp} shorter than 15 years\footnote{The investment decisions are optimized by maximizing the net present value over the 30-year lifetime of the \gls{pvb} system, which is not equivalent to allow all investments that have a \gls{pvb} below 30 years. This is because the \gls{npv} is calculated considering the time value of the money, which is not the case when calculating the \gls{pbp}.}, which can also be seen in Fig.~\ref{fig:Baseline_PBPdistribution_ZH}. 
In 2050, almost all investments achieve a \gls{pbp} of less than 10 years.

\begin{figure}[h!]
\vspace{-0.2cm}
  \centering
    \begin{subfigure}[b]{0.51\textwidth}
      \includegraphics[width=\textwidth]{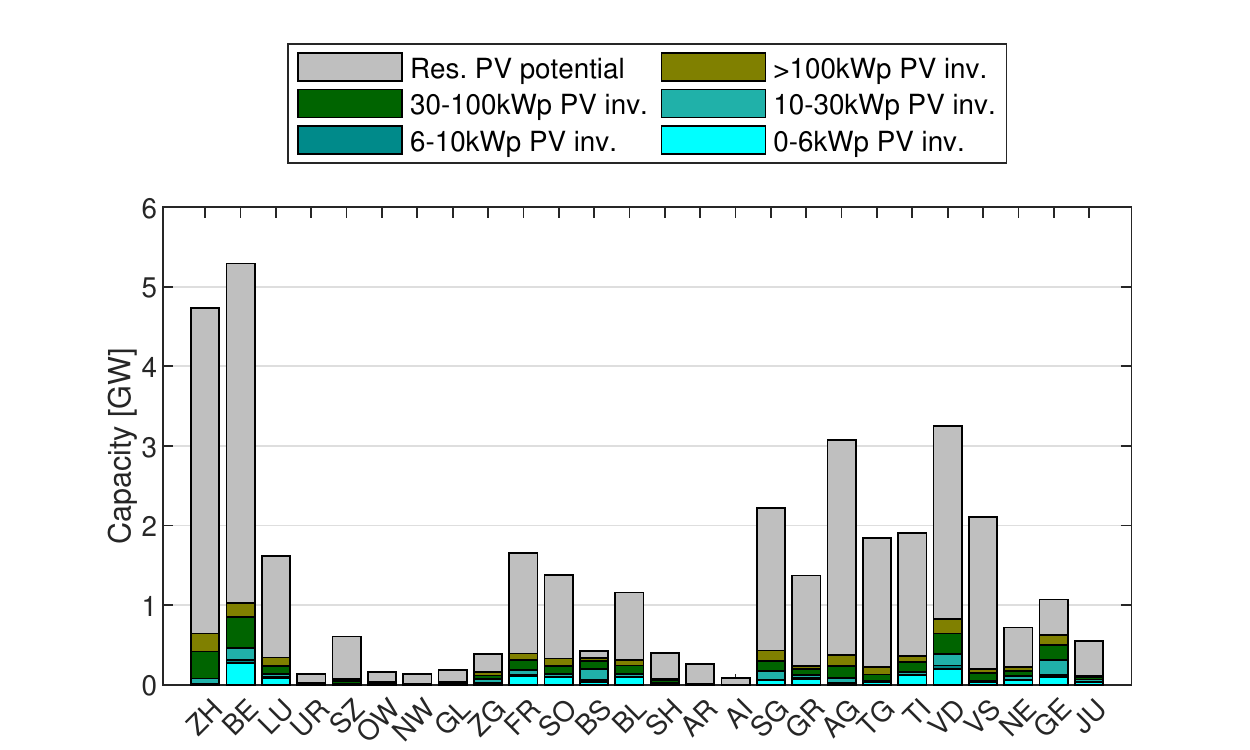}
      \caption{Fast recoverable investment - 2020}
      \label{fig_Baseline_PBP10year_regional_2020}
    \end{subfigure}
\vfill
    \begin{subfigure}[b]{0.505\textwidth}
      \includegraphics[width=\textwidth]{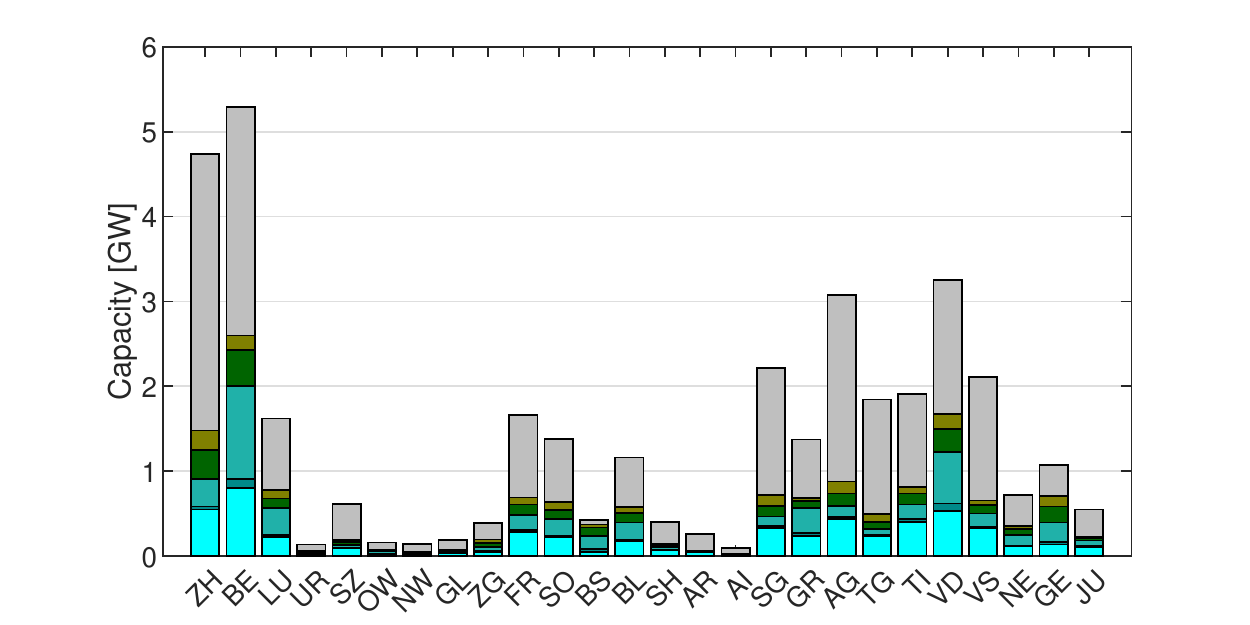}
      \caption{Moderately fast recoverable investment - 2020}      \label{fig_Baseline_PBP15year_regional_2020}
    \end{subfigure}
\vfill
    \begin{subfigure}[b]{0.505\textwidth}
      \includegraphics[width=\textwidth]{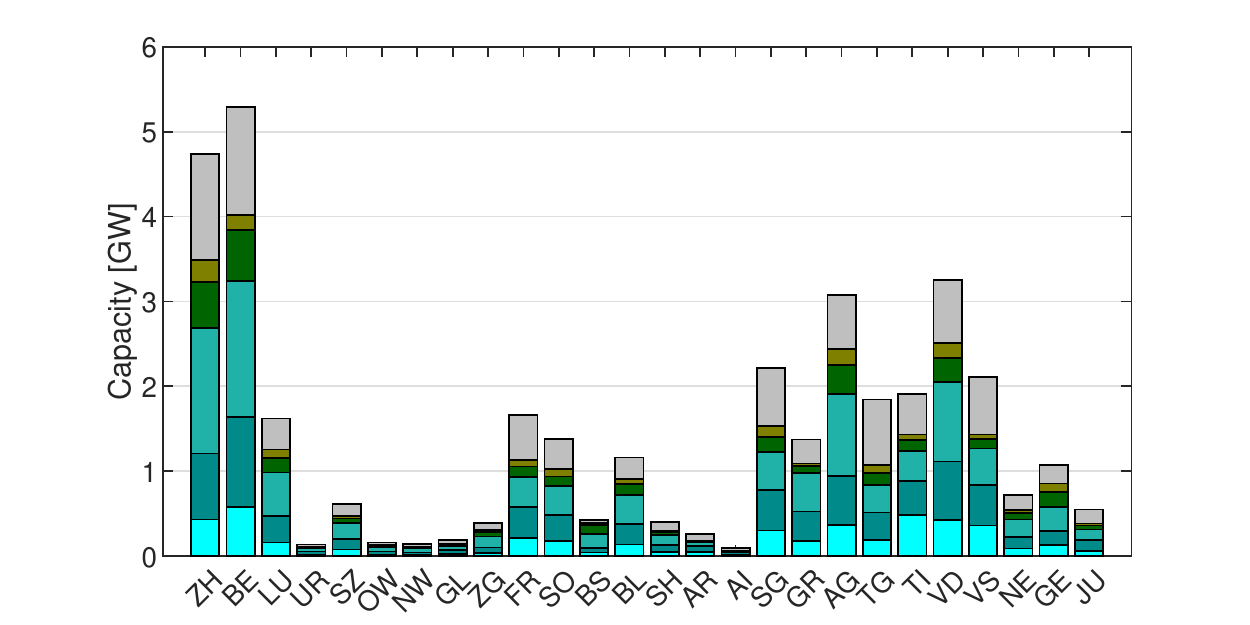}
      \caption{Fast recoverable investment - 2050}      \label{fig_Baseline_PBP10year_regional_2050}
    \end{subfigure}
\vfill
    \begin{subfigure}[b]{0.505\textwidth}
      \includegraphics[width=\textwidth]{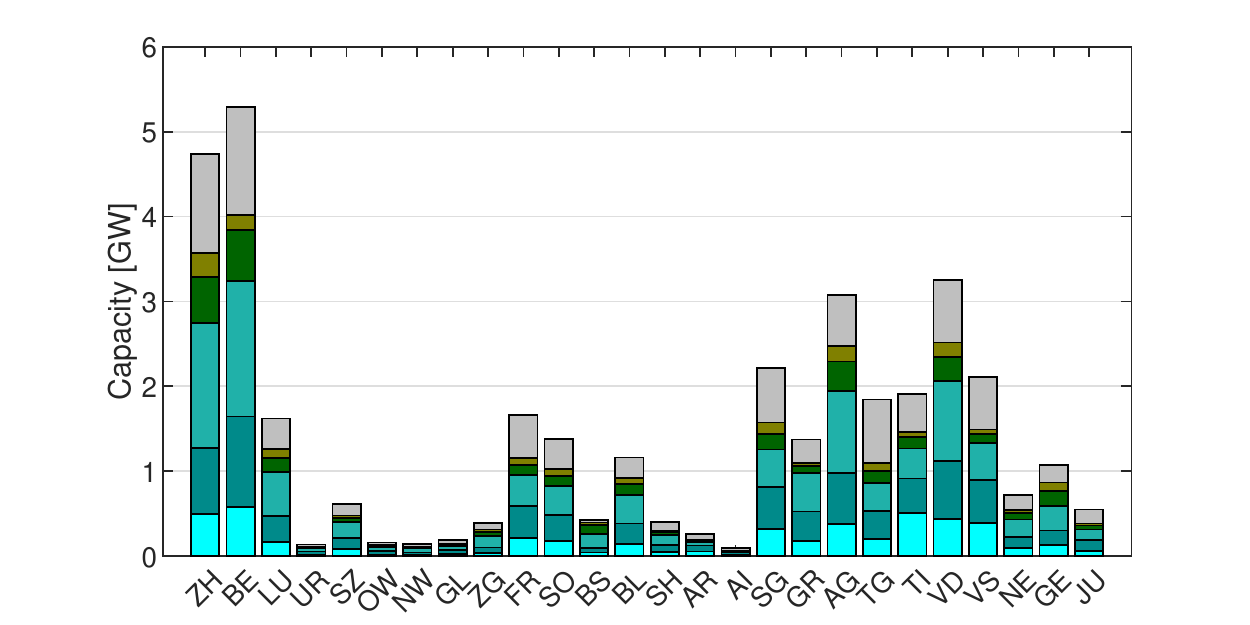}
      \caption{Moderately fast recoverable investment - 2050}      \label{fig_Baseline_PBP15year_regional_2050}
    \end{subfigure}
    \caption{Optimal regional investment of a fast and a moderately fast recoverable investment cases of the Baseline scenario in 2020 and 2050.}\label{fig:Baseline_regional}
\vspace{-0.2cm}
\end{figure}

Figure~\ref{fig:Baseline_regional} shows the regional investment capacities of fast recoverable and moderately fast recoverable investments in both 2020 and 2050, and \gls{pv} investments are broken down into different \gls{pv} size categories. In 2020, the fast recoverable investments are mainly large \gls{pv} units. In cantons with high \gls{dso} injection tariffs (e.g. BS and GE), a significant share of deployment potentials is already qualified as fast recoverable in 2020. While in 2050, the fast recoverable investments are more evenly distributed between different regions and different \gls{pv} categories.
Moreover, profitable \gls{pv} investment capacities increase while the corresponding \glspl{pbp} decrease from 2020 to 2050.

\begin{figure}[t]
\vspace{-0.2cm}
  \centering
    \begin{subfigure}[b]{0.52\textwidth}
      \includegraphics[width=\textwidth]{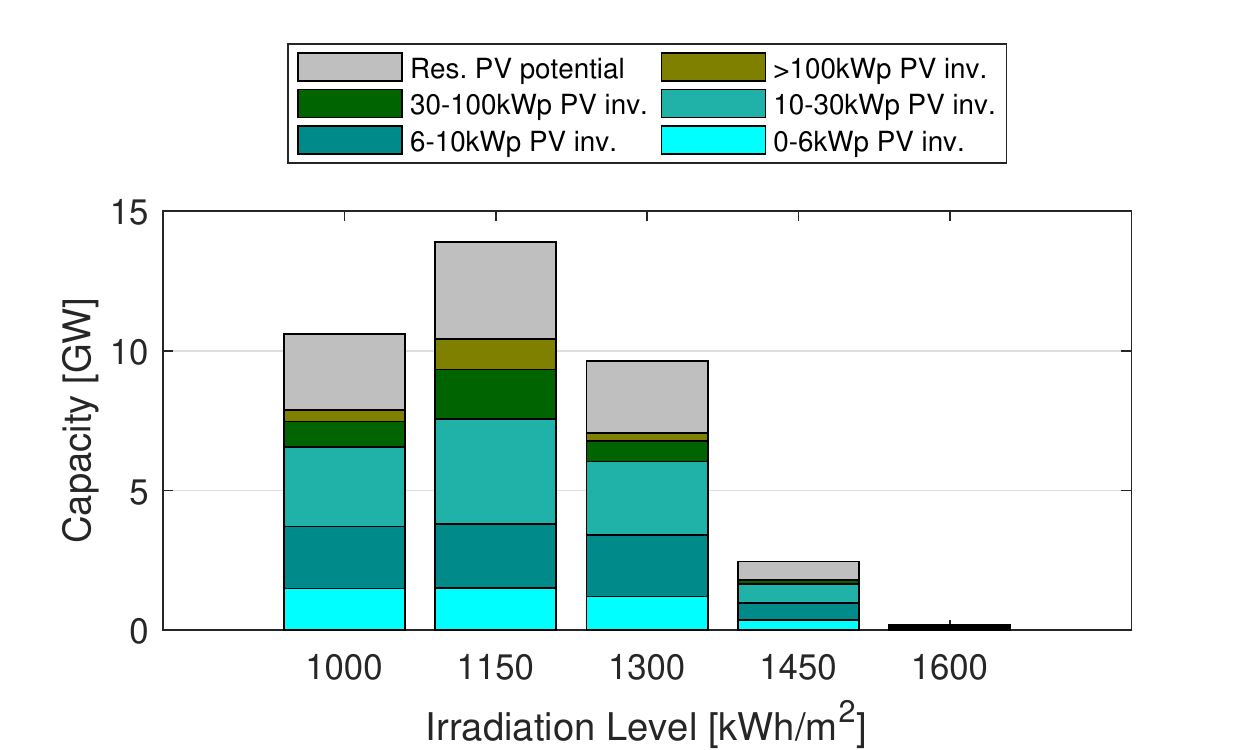}
      \caption{Annual irradiation - PV investment}
      \label{fig_Baseline_irr_2050}
    \end{subfigure}
        \begin{subfigure}[b]{0.52\textwidth}
      \includegraphics[width=\textwidth]{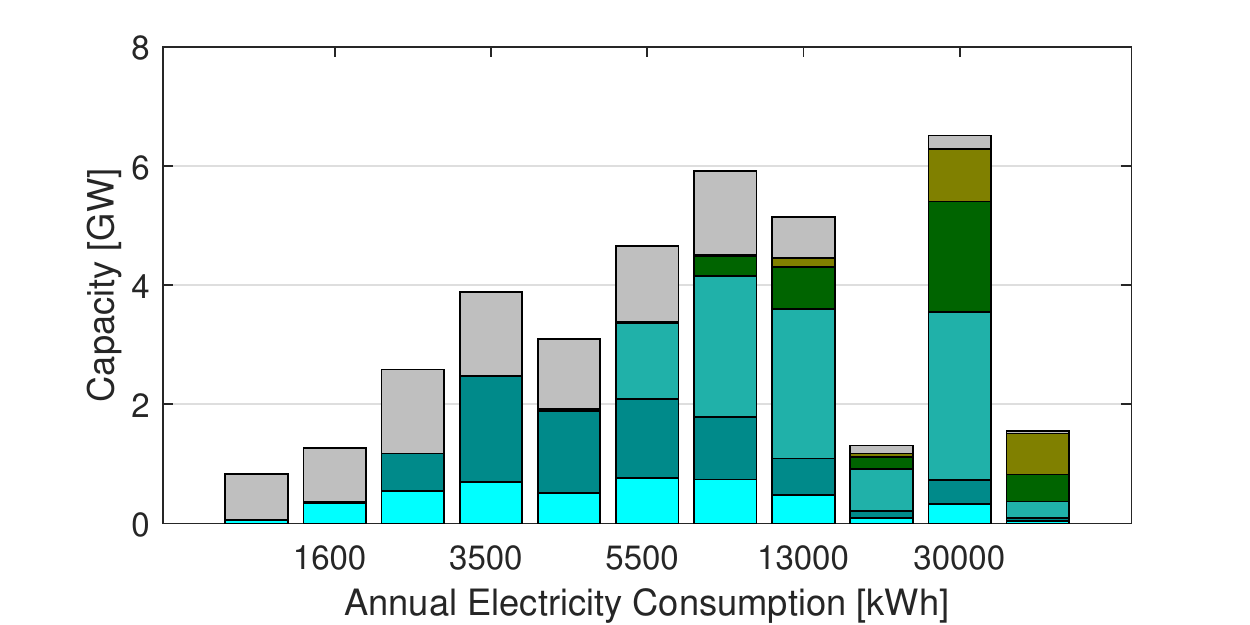}
      \caption{Electricity consumption - PV investment}
      \label{fig_Baseline_consumption_2050}
    \end{subfigure}
    \begin{subfigure}[b]{0.52\textwidth}
      \includegraphics[width=\textwidth]{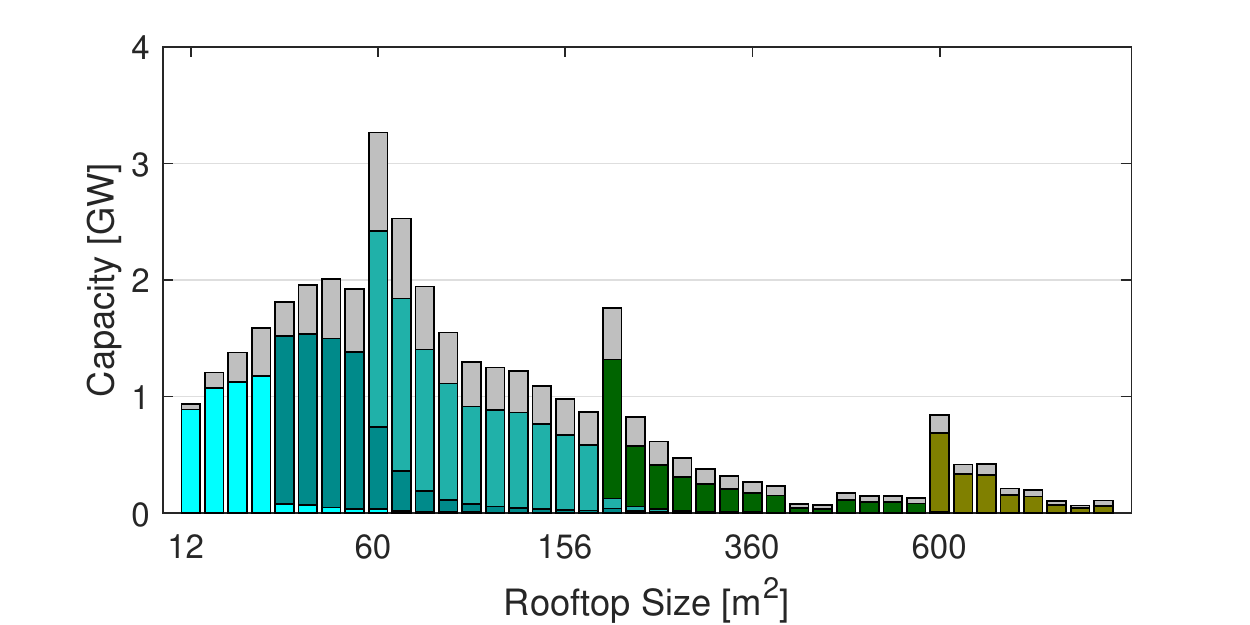}
      \caption{Rooftop area - PV investment}
      \label{fig_Baseline_roofarea_2050}
    \end{subfigure}
    \caption{Distribution of the fast recoverable investment of the Baseline scenario in 2050.}\label{fig:Baseline_levels_2050}
\vspace{-0.2cm}
\end{figure}

Figure~\ref{fig:Baseline_levels_2050} presents distributions of the fast recoverable investments in 2050 over different irradiation, rooftop size and annual electricity consumption categories. It can be noticed in Fig. \ref{fig_Baseline_irr_2050} and Fig. \ref{fig_Baseline_consumption_2050} that the most attractive investments mainly belong to the customer groups that are in the higher annual irradiation and higher electricity consumption categories. Furthermore, the optimal \gls{pv} investment size is generally limited by the rooftop size, as illustrated in Fig.~\ref{fig_Baseline_roofarea_2050} with the separated ordering of the colored \gls{pv} categories from light to dark green, which shows the importance of considering rooftop size limits in the techno-economic model.

\subsubsection{Baseline results - load}
\begin{figure}[t]
  \centering
    \begin{subfigure}[b]{0.5\textwidth}
      \includegraphics[width=\textwidth]{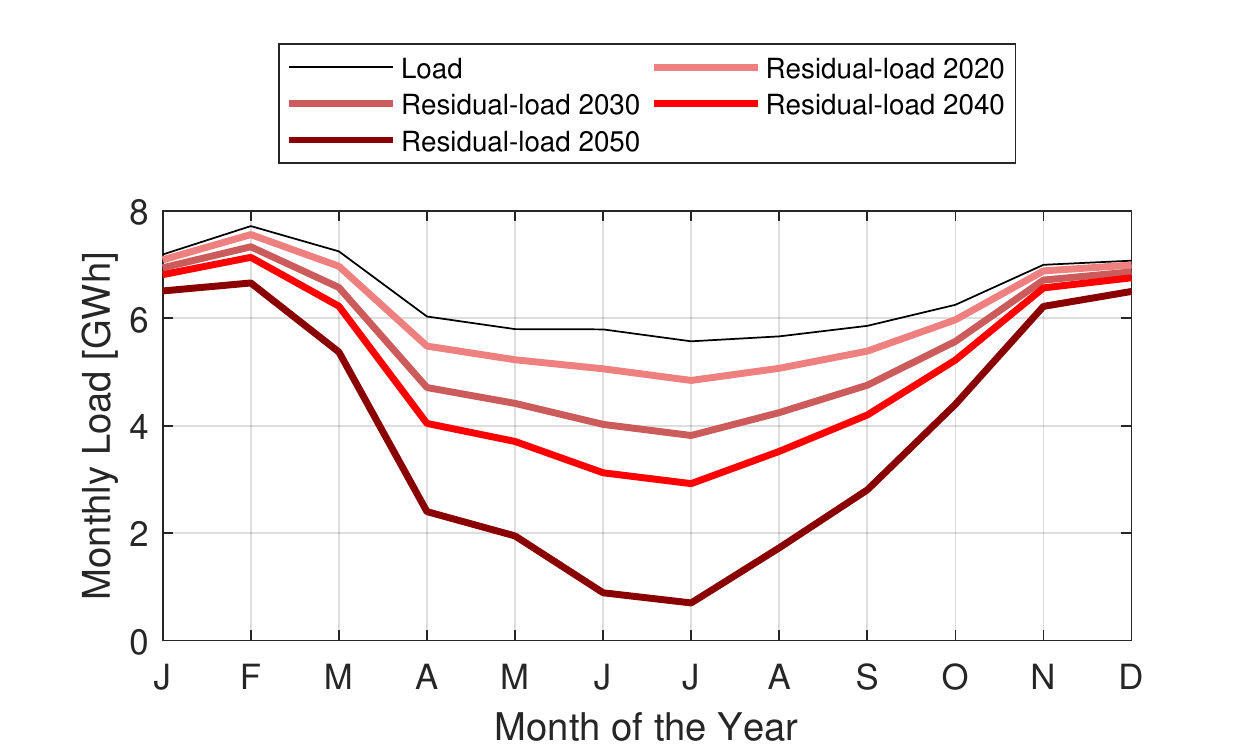}
      \caption{Fast recoverable investment case}
      \label{fig_CHdispatch_monthly_10PBP}
    \end{subfigure}
    \begin{subfigure}[b]{0.5\textwidth}
      \includegraphics[width=\textwidth]{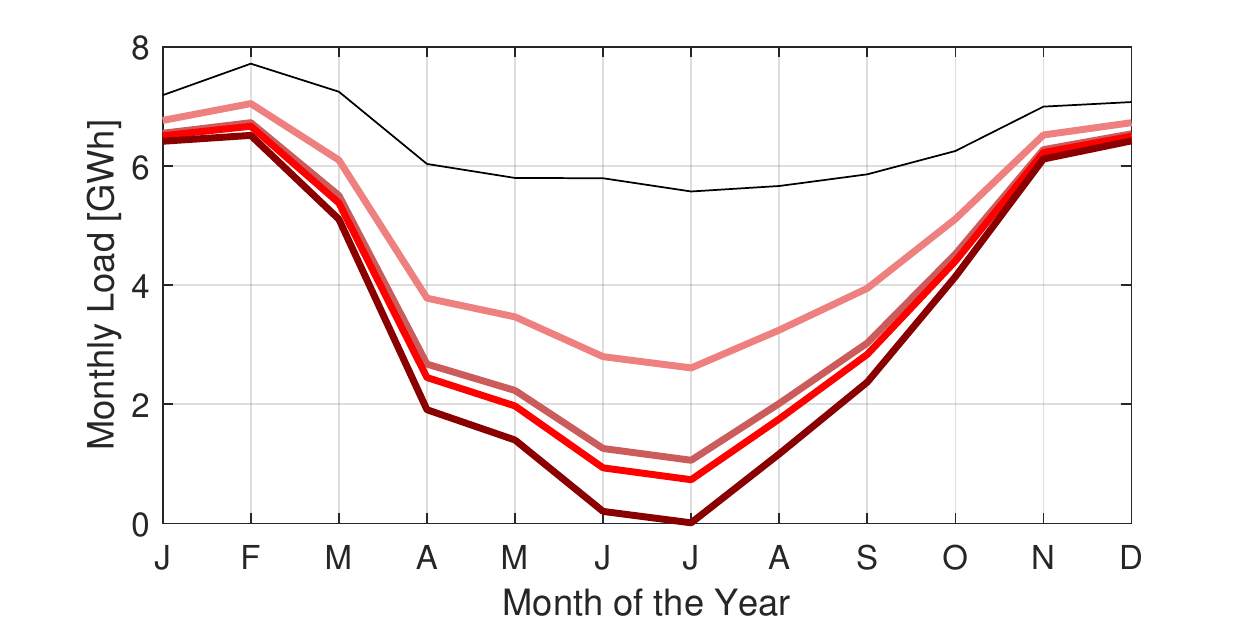}
      \caption{Moderately fast recoverable investment case}      \label{fig_CHdispatch_monthly_15PBP}
    \end{subfigure}
    \caption{Monthly load and residual-load of different years under a fast and a moderately fast recoverable investment case.}\label{fig:CHdispatch_monthly}
\end{figure}
Figure~\ref{fig:CHdispatch_monthly} shows the average hourly Swiss load per month and the average hourly residual load between 2020-2050 assuming either a fast or a moderately fast recoverable investment case, i.e. investments are only made if the \gls{pbp} is less than 10 years or less than 15 years, respectively. This residual load represents the Swiss demand that is not supplied by the invested \gls{pvb} units, which is equal to the Swiss load minus the consumers' load that is self-supplied by the invested \gls{pvb} units and minus the excess \gls{pv} generation that is injected into the grid.  
The original Swiss load profile is shown to be higher in winter months, with a peak in January, and lower in summer months. This seasonal pattern is impacted by higher electricity demand during the cold winter months along with very limited existing cooling in summer. We see that while the residual load remains high in winter of all years, it is increasingly reduced over time during summer, which is directly attributable to the cumulative \gls{pv} installations over all customer groups and regions that fulfill the respective \gls{pbp} limit in the considered year. 
This seasonality is more pronounced when moving from fast recoverable to moderately fast recoverable investments since the less stringent \gls{pbp} threshold enables more \gls{pv} to be viable. It is worth noting that the original Swiss load is assumed to be constant over the years, but expectations for future demand changes as well as electrification are not expected to make significant differences to the seasonal pattern of the Swiss demand.

\begin{figure}[h!]
\vspace{-0.2cm}
  \centering
    \begin{subfigure}[b]{0.49\textwidth}
      \includegraphics[width=\textwidth]{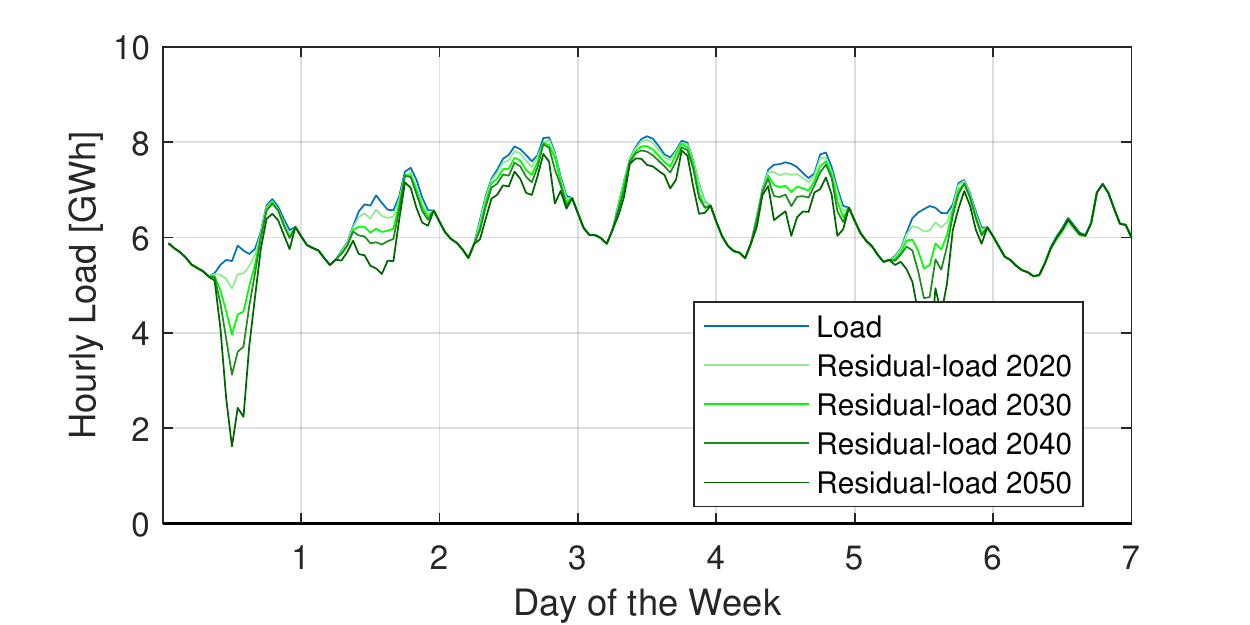}
      \caption{Winter week in fast recoverable investment case}
      \label{fig_CHdispatch_winter_10PBP}
    \end{subfigure}
    \vfill
    \begin{subfigure}[b]{0.49\textwidth} 
      \includegraphics[width=\textwidth]{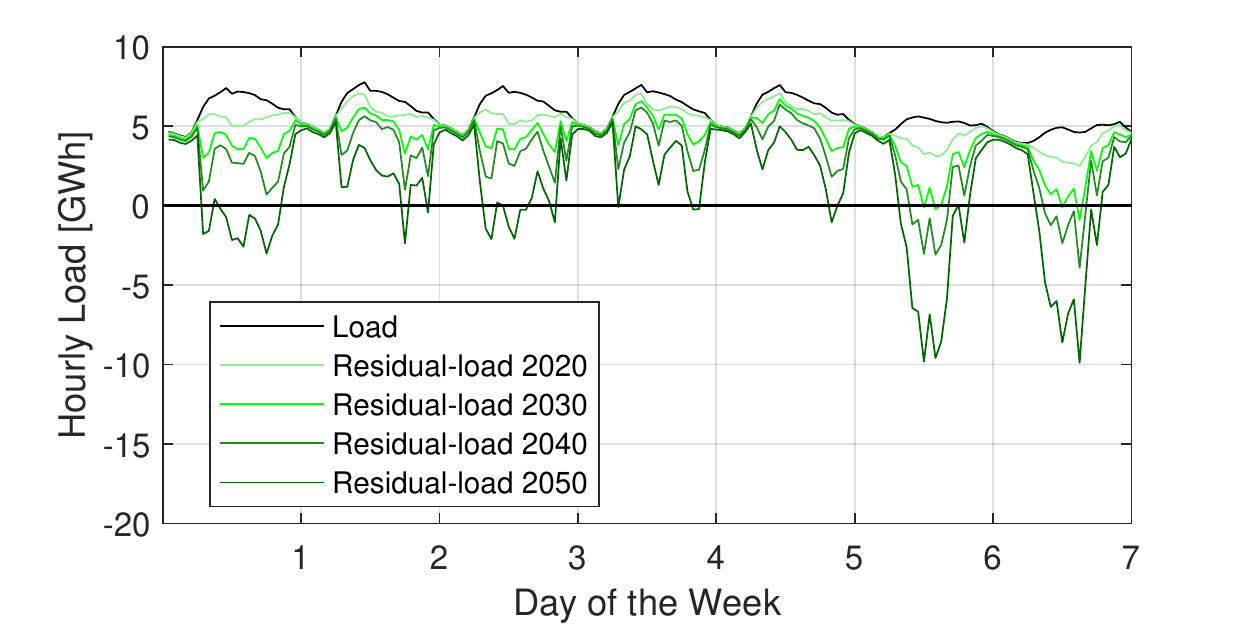}
      \caption{Summer week in fast recoverable investment case}
      \label{fig_CHdispatch_summer_10PBP}
    \end{subfigure}
    \vfill
        \begin{subfigure}[b]{0.49\textwidth}
      \includegraphics[width=\textwidth]{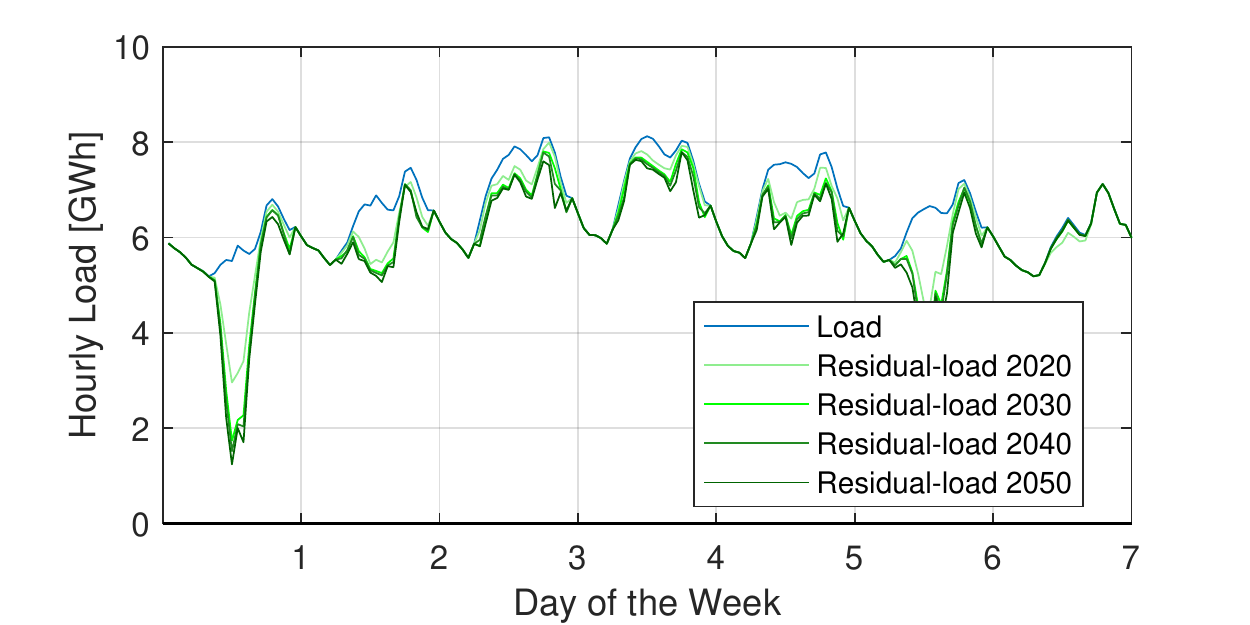}
      \caption{Winter week in moderately fast recoverable investment case}
      \label{fig_CHdispatch_winter_15PBP}
    \end{subfigure}
    \vfill
        \begin{subfigure}[b]{0.49\textwidth}
      \includegraphics[width=\textwidth]{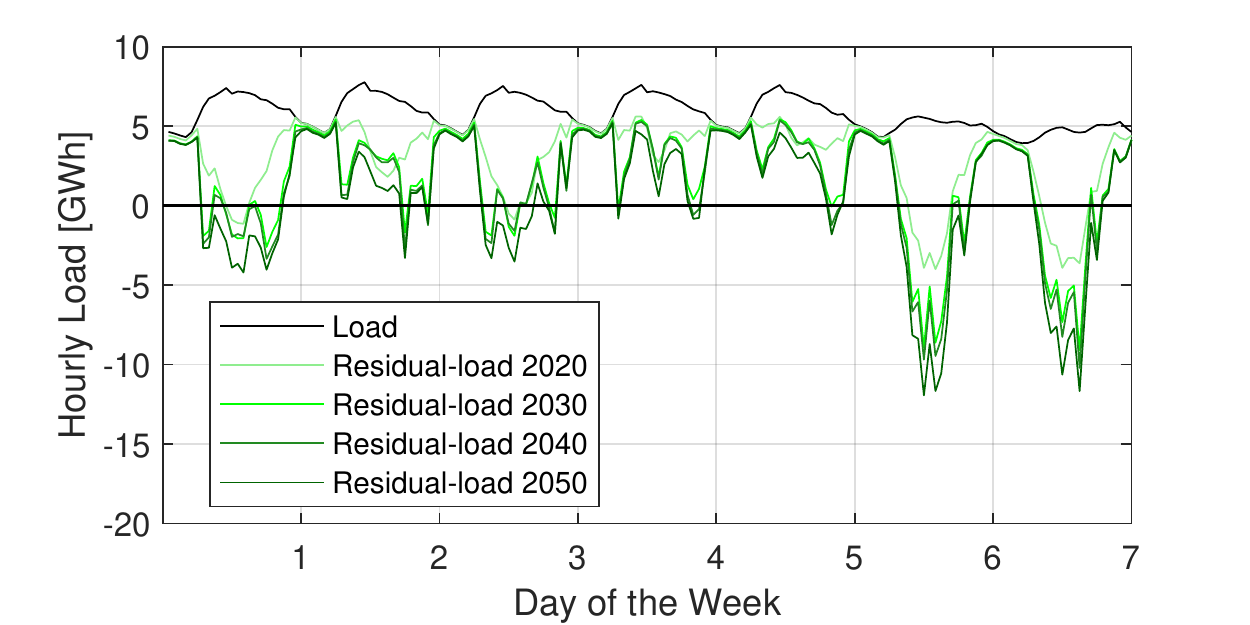}
      \caption{Summer week in moderately fast recoverable investment case}
      \label{fig_CHdispatch_summer_15PBP}
    \end{subfigure}
    \caption{Hourly original Swiss load and residual-load of a winter and a summer week under a fast and a moderately fast recoverable investment case.}\label{fig:CHdispatch_hourly}
\vspace{-0.4cm}
\end{figure}

Figure~\ref{fig:CHdispatch_hourly} shows the original hourly load of Switzerland and the residual load for the years 2020 through 2050 for one winter week in January and one summer week in July for both the fast recoverable and the moderate fast recoverable investment cases. 
In all weeks, the original Swiss load follows a similar pattern with higher consumption during the day and less at night. Over the years, in general the residual load profiles deviate more and more from the original Swiss load profile during the afternoon hours when the \gls{pv} generation peaks. 
One exception can be observed on the 7th day (i.e., Sunday) in Fig.~\ref{fig_CHdispatch_winter_10PBP} and Fig.~\ref{fig_CHdispatch_winter_15PBP} when the residual load profile in 2020 is lower than or similar to that of 2030-2050. This is due to the fact that although \gls{pv} generation is higher in 2030-2050, more batteries are also installed in 2030-2050 and absorb the \gls{pv} generation during these hours, which is relatively lower due to the winter season, while the load is instead supplied by the grid at the low retail electricity tariff available during these Sunday hours.  
By 2050, every sunny day in both the winter and summer weeks exhibits a highly dynamic plunge and recovery pattern. It can be seen that the residual load can vary drastically from one day to the next and also from one hour to the next. Both phenomena become more pronounced as the \gls{pv} penetration level increases from 2020 to 2050 and the analyzed investments extend from fast recoverable to the moderately fast recoverable ones. 
The increasingly dynamic pattern of the residual load on an hourly and daily basis emphasizes the need for flexible resources with fast ramping capabilities.
%
\subsubsection{Baseline results - self-consumption}
Table~\ref{tab:selfconsumption} shows the Baseline self-consumption results analysis for the fast and moderately fast recoverable investments from 2020 to 2050.
\begin{table}[t]
\renewcommand{\arraystretch}{1.1}
\caption{Baseline self-consumption results analysis for the fast / moderately fast recoverable investments, years 2020-2050.}
\label{tab:selfconsumption}
\centering
\begin{tabular}{|l|c|c|c|c|}
\hline
\textbf{Year} & \textbf{\begin{tabular}[c]{@{}c@{}}PV \\ generation\\ {[}TWh{]}\end{tabular}} & \textbf{SCR} & \textbf{\begin{tabular}[c]{@{}c@{}}Savings\\ {[}bn. EURs{]}\end{tabular}} & \textbf{\begin{tabular}[c]{@{}c@{}}Required retail \\ tariff increase \\ {[}cent/kWh{]}\end{tabular}} \\ \hline
\textbf{2020} & 3.5 / 14.3 & 54\% / 64\% & 0.4 / 1.8 & 0.7 / 4.3 \\ \hline
\textbf{2030} & 8.7 / 22.2 & 74\% / 72\% & 1.4 / 3.4 & 3.0 / 9.7 \\ \hline
\textbf{2040} & 13.1 / 23.9 & 76\% / 73\% & 2.4 / 4.0 & 5.6 / 12.0 \\ \hline
\textbf{2050} & 24.0 / 27.3 & 68\% / 67\% & 4.2 / 4.7 & 12.6 / 15.4 \\ \hline
\end{tabular}
\end{table}
It can be seen that in both cases while the \gls{pv} generation increases over time, the \gls{scr} peaks in 2040 since later investments are more driven by the low cost of the \gls{pvb} system than by trying to increase the \gls{scr}.
We can now also analyze the inherent losses for the retailers and \glspl{dso} caused by the reduced electricity purchase and compute by how much they would have to increase the retail price in order to recover these losses. 
The extra retail electricity tariff charge is calculated as the revenue loss of the \glspl{dso} divided by the sum of the residual load. The revenue loss is assumed to be equal to the savings earned by end-consumers on their electricity bills as a result of self-consumed \gls{pv} generation instead of purchasing from the grid. 
From Table~\ref{tab:selfconsumption} we can see that this extra required retail electricity tariff calculated rises significantly over the years, especially from 2040 to 2050, which is due to the strong increase in self-consumption savings and the reduction in residual loads. Using the Swiss average household tariff in 2020 (i.e., 18.8 cent/kWh \cite{web_globalpetrolprice}) as a reference, the 12.6 cent/kWh and the 15.4 cent/kWh required tariff increase in 2050 in the fast and moderately fast recoverable investment cases translate into a total increase of 67\% and 82\%, which is equivalent to a yearly increase of 1.7\% and 2.0\% between 2020-2050. This of course is a simplified analysis as it does not take into account the rebound effect on \gls{pv} and battery investments that would be driven by the increase in these retail prices but it points to an important issue that retailers and \glspl{dso} will likely face in the future.


\section{Discussions}
\label{sec:Discussions}





\subsection{From investors' perspective}
Results show that combining a battery with a \gls{pv} unit is in some cases already economically viable today and especially for investors that have high annual electricity consumption. However, the average payback period 
might fluctuate between 2020 and 2035 mainly due to the mixed impacts of subsidy policy changes, cost reductions, injection tariff and electricity price developments.
A significant decrease of the \gls{pbp} is expected after 2040.

In addition, profitability and optimal size of the \gls{pvb} system vary among consumer groups due to their diverse annual electricity consumption, locations, solar irradiation and rooftop sizes. It is therefore important to consider the heterogeneity of different investors when assessing the economic viability of the investment.

Furthermore, the economics of the \gls{pvb} system are especially sensitive to \gls{pv} and battery cost developments, injection tariff changes and wholesale and retail electricity price changes. Having access to the wholesale market can either increase or decrease the economic viability of the \gls{pvb} system, depending on how the retail and wholesale electricity prices develop in the future and their relationship to each other. 

Please note that the assumptions made in this work heavily affect our results obtained from the scenario simulations and we do not account for idealistic motivations, i.e., non-economical reasons, to install \gls{pv} and battery units. Therefore, we do not claim that the optimized investment decisions will be realized, given the modeled regulatory and legislative framework. However, the results indicate how the development of costs and electricity prices over the years affect the \gls{pvb} system investment and operation decisions. Also, we conducted sensitivity analyses to better understand how our assumptions on different parameters (e.g., payback period and unit costs) affect the potential investments in \gls{pvb} systems.

\subsection{From retailers' perspective}
According to our results with respect to the fast and moderately fast recoverable investments, by the end of 2050, 24.0 TWh and 27.3 TWh of the \gls{pv} generation, which account for 37.9\% and 43.1\% of the total Swiss demand in 2019 (i.e. 63.4 TWh), could be self-consumed by the end-consumers. The resulting revenue losses from the decrease in electricity purchases of the prosumers could be recovered by increasing the retail electricity tariff. However, a higher retail electricity tariff will in turn further encourage investments in \gls{pv} and battery units.

The current retail electricity tariff (including the grid tariff) scheme is mainly energy-based (i.e. electricity charged based on the kWh of electricity consumed). Although the \gls{pvb} system investments tend to decrease the annual net electricity consumption, the possibly higher dynamics of their residual load profile and the absolute value of their peak net-load will increase the burden on the grid. Therefore, an additional capacity-based grid tariff could enable a more reasonable pricing scheme and incentivize the prosumers to optimize their dispatch more in favor of the grid.  

However, it is also worth mentioning that the future load profile is also subject to uncertainties brought by the electrification of heating and the transportation sector, which could compensate the loss of the self-consumption or further exacerbate the problem by increasing the self-consumption.

\subsection{From system operators' perspective}
The seasonality of the residual load increases in all future years and is more pronounced in 2050 as the \gls{pv} penetration level is higher. This trend indicates a need for greater seasonal flexibility that could impact the operation of centralized power plants as well as the pattern of imports and exports.
Similarly, the increasingly dynamic pattern of the residual load on an hourly and daily basis emphasizes the need for flexible resources with fast ramping capabilities such as hydro dams with storage reservoirs and load shifting units like pumped hydro, battery storage, or \gls{dsm}. Furthermore, in future years, more frequent curtailments of non-dispatchable units, while in general not being popular, will likely make economic sense. 
With the integration of high levels of non-dispatchable \gls{pv}, a shortage of flexibility in the power system will have a negative effect on grid security and, thus, will contribute to the risk of systemic failures.
All these issues will potentially impact the actions of and services provided by the system operator.

\subsection{From policy-makers' perspective}
The optimization results in the Baseline scenario show that most \gls{pvb} systems could result in positive \gls{npv} even without policy support after 2030. However, the payback periods of the invested projects fluctuate between 2020 and 2040 in the Baseline scenario, which indicates that there are competing influences that could reduce the economic attractiveness for customers.
The increasing dynamics of residual load profiles require increased levels of flexibility provisions in the distribution and transmission networks while at the same time lower \gls{dso} revenues are expected as a result of lower electricity purchases of prosumers.
Policy-makers therefore may have to rethink the market design and rules to not only promote investments in renewable generations but also in resources that are capable of providing the required flexibility. 
We believe that the sensitivity analyses in this work that assess the impacts of different input parameters enable policy-makers to identify the main driving factors for investments in \gls{pv} and battery units.



\section{Limitations and future work}
\label{sec:Limitation and future work}
This work has several limitations and a few of which are highlighted in this section. First, we estimate the load profiles of different customer groups using a number of synthetic load profiles that undoubtedly deviate from real-world data. Thus, we do not capture the variety of different consumption behaviors between various regions and sectors. Furthermore, due to the lack of input data, we approximate the annual electricity consumption of individual customers using the warm water consumption data; additionally, the annual electricity consumption value is assumed to be constant over the years, which does not capture the possibility of an increase in \gls{ev} penetrations or other electrification. Future work should include a bottom-up representation of buildings’ electricity demand by utilizing realistic load patterns that evolve from year to year and differ from region to region. 

Second, we group the rooftop data using a limited number of clusters and represent each group using the median data. The groups with highly varied data and the groups with few data therefore cannot be well represented; this is especially important since these two cases mostly correspond to the groups with high electricity consumption or large roofs. However, increasing the number of clusters causes higher computational complexity and longer simulation time. In future work, a proper clustering method possibly is required and a comprehensive analysis is needed to investigate the impact of the clustering on the results. 

Third, we model the investment behavior using a \gls{npv}-maximization objective without considering non-economic factors. 
A future version should account for a heterogeneous investor population, including, for example, varying risk profiles and cost-unrelated objectives such as peer-effects. These enhancements toward a diverse consumer perspective would enable a more realistic assessment of the investment decisions and their impacts on the grid. However, completing such improvements requires additional input data and increases the implementation overhead.

Fourth, we adopt historical wholesale market prices and apply different multiplication factors to simulate the different wholesale electricity price scenarios in the future. However, in this way the price suppression effect of \gls{pvb} system injections especially during high \gls{pv} generation hours cannot be captured. In addition, because of the central hub position of Switzerland, the Swiss wholesale electricity prices are also impacted by generation mix changes in surrounding countries.

And finally, a higher time resolution could be employed for both the hourly operational decisions and the 5-year investment optimization to enable a more granular assessment of operations and the economics for investing in \gls{pvb} systems.


\section{Conclusions}
\label{sec:Conclusions}

This paper presents a techno-economic optimization model to analyze the economic viability of \gls{pvb} systems for different customer groups in Switzerland clustered based on their annual electricity consumption, rooftop size, annual irradiation, and region. There are in total 2200 customer groups considered for each of the 26 regions in Switzerland. Each of the customer groups is represented using median values for each of the dimensions that define the group. The optimization of a static investment model is carried out considering a greenfield investment for each of the investment years from 2020 through 2050 (i.e., each year run independently without taking investments from previous years into account). The resulting optimal decisions are then applied to all customers within the corresponding customer group. 
A comprehensive sensitivity analysis is conducted for an example of a particular canton (i.e., Zurich) in 2050 to investigate the impacts of input parameters such as costs, load profiles, electricity prices and tariffs on the optimal investment decisions. 

Results show that the combined \gls{pv} plus battery system investments for some customer groups already yield a better \gls{npv} than \gls{pv} alone today. 
The payback period of \gls{pvb} system investments fluctuates between 2020 and 2035 due to the mixed effects of policy changes, costs, and electricity price developments, but decreases significantly afterwards.
The optimal \gls{pv} and battery sizes increase over time. In 2050, the \gls{pvb} system investment is profitable for most customer groups and the \gls{pv} investment with the shortest \gls{pbp} is mostly limited by the rooftop size. 
Optimal investment decisions vary between different customer groups and fast recoverable investment (i.e., with the shortest \gls{pbp}) is mostly accessible to customer groups that have high annual irradiation and electricity demand, which suggests that it is important to consider the heterogeneity of different customer groups when assessing the economic viability of \gls{pvb} system investments.
With regard to the grid impact, dynamics of new system load profiles caused by the seasonal, daily and hourly patterns of the solar generation emphasize the need for system flexibility. 
Furthermore, the electricity purchases of the end-consumers decrease dramatically over the years since more consumers turn into prosumers. Such a change could require rethinking the current electricity tariff and subsidy policy design.
In addition, investment decisions are highly sensitive to the expected payback periods, future costs, injection tariff developments, and wholesale and retail electricity price changes. It is therefore important to identify the driving  factors of the \gls{pvb} system investments and understand the future uncertainties of different input parameters when discussing the economic viability of \gls{pvb} systems in the future.

\newpage
\clearpage

\bibliographystyle{unsrt}

\raggedright
\bibliography{ms}

\newpage
\clearpage

\begin{appendices}
\section{Input data of a low, baseline and high cost scenario for PV units}
\label{Apdx:App-futurecosts_pv}


\begin{table}[hbtp]
\renewcommand{\arraystretch}{1.2}
\caption{Baseline PV cost scenario for 2020-2050 \cite{bauer2017potential,Bauer2019}.}
\label{tab:cost_scen_ref_PV}
\centering
\begin{tabular}{|l|c|c|c|c|c|}
\hline
 &  & \textbf{2020} & \textbf{2030} & \textbf{2040} & \textbf{2050} \\ \hline
\multirow{5}{*}{\textbf{\begin{tabular}[c]{@{}l@{}}PV investment \\ cost (\euro/kWp)\end{tabular}}} & \textbf{0-6 kWp} & 2'496 & 2'060 & 1'770 & 1'654 \\ \cline{2-6} 
 & \textbf{6-10 kWp} & 2'393 & 1'964 & 1'578 & 1'204 \\ \cline{2-6} 
 & \textbf{10-30 kWp} & 1'916 & 1'572 & 1'308 & 1'102 \\ \cline{2-6} 
 & \textbf{30-100 kWp} & 1'272 & 1'036 & 895 & 816 \\ \cline{2-6} 
 & \textbf{\textgreater{}100 kWp} & 814 & 664 & 573 & 523 \\ \hline
\multirow{5}{*}{\textbf{\begin{tabular}[c]{@{}l@{}}PV operational\\ cost (cent/kWh)\end{tabular}}} & \textbf{0-6 kWp} & 2.6 & 2.1 & 1.9 & 1.7 \\ \cline{2-6} 
 & \textbf{6-10 kWp} & 2.6 & 2.1 & 1.8 & 1.7 \\ \cline{2-6} 
 & \textbf{10-30 kWp} & 2.6 & 2.1 & 1.8 & 1.7 \\ \cline{2-6} 
 & \textbf{30-100 kWp}  & 2.6 & 2.1 & 1.8 & 1.7 \\ \cline{2-6}
 & \textbf{\textgreater{}100 kWp}  & 1.7 & 1.4 & 1.2 & 1.2 \\ \hline
\end{tabular}
\end{table}

\begin{table}[htbp]
\renewcommand{\arraystretch}{1.2}
\caption{High cost scenario for 2020-2050 \cite{bauer2017potential,Bauer2019}.}
\label{tab:cost_scen_high_PV}
\centering
\begin{tabular}{|l|c|c|c|c|c|}
\hline
 &  & \textbf{2020} & \textbf{2030} & \textbf{2040} & \textbf{2050} \\ \hline
\multirow{5}{*}{\textbf{\begin{tabular}[c]{@{}l@{}}PV investment \\ cost  (\euro/kWp)\end{tabular}}} & \textbf{0-6 kWp} & 2'786 & 2'322 & 2'060 & 1'857 \\ \cline{2-6} 
 & \textbf{6-10 kWp} & 2'546 & 2'241 & 1'854 & 1'411 \\ \cline{2-6} 
 & \textbf{10-30 kWp} & 2'066 & 1'813 & 1'561 & 1'308 \\ \cline{2-6} 
 & \textbf{30-100 kWp} & 1'382 & 1'225 & 1'083 & 989 \\ \cline{2-6} 
 & \textbf{\textgreater{}100 kWp} & 885 & 784 & 694 & 633 \\ \hline
\multirow{5}{*}{\textbf{\begin{tabular}[c]{@{}l@{}}PV operational\\ cost (cent/kWh)\end{tabular}}} & \textbf{0-6 kWp} & 2.6 & 2.2 & 1.9 & 1.7 \\ \cline{2-6} 
 & \textbf{6-10 kWp} & 2.6 & 2.2 & 1.9 & 1.7 \\ \cline{2-6} 
 & \textbf{10-30 kWp} & 2.6 & 2.2 & 1.9 & 1.7 \\ \cline{2-6} 
 & \textbf{30-100 kWp}  & 2.6 & 2.2 & 1.9 & 1.7 \\ \cline{2-6}
 & \textbf{\textgreater{}100 kWp}  & 1.7 & 1.5 & 1.3 & 1.2 \\ \hline
\end{tabular}
\end{table}

\begin{table}[htbp]
\renewcommand{\arraystretch}{1.2}
\caption{Low PV cost scenario for 2020-2050 \cite{bauer2017potential,Bauer2019}.}
\label{tab:cost_scen_low_PV}
\centering
\begin{tabular}{|l|c|c|c|c|c|}
\hline
 &  & \textbf{2020} & \textbf{2030} & \textbf{2040} & \textbf{2050} \\ \hline
\multirow{6}{*}{\textbf{\begin{tabular}[c]{@{}l@{}}PV investment \\ cost  (\euro/kWp)\end{tabular}}} & \textbf{0-6 kWp} & 2'351 & 1'799 & 1'480 & 1'422 \\ \cline{2-6} 
 & \textbf{6-10 kWp} & 2'241 & 1'715 & 1'300 & 996 \\ \cline{2-6} 
 & \textbf{10-30 kWp} & 1'790 & 1'354 & 1'056 & 996 \\ \cline{2-6} 
 & \textbf{30-100 kWp} & 1'178 & 864 & 691 & 644 \\ \cline{2-6} 
 & \textbf{\textgreater{}100 kWp} & 754 & 553 & 442 & 412 \\ \hline
\multirow{6}{*}{\textbf{\begin{tabular}[c]{@{}l@{}}PV operational\\ cost  (cent/kWh)\end{tabular}}} & \textbf{0-6 kWp} & 2.6 & 2.1 & 1.9 & 1.7 \\ \cline{2-6} 
 & \textbf{6-10 kWp} & 2.6 & 2.1 & 1.8 & 1.7 \\ \cline{2-6} 
 & \textbf{10-30 kWp} & 2.6 & 2.1 & 1.8 & 1.7 \\ \cline{2-6} 
 & \textbf{30-100 kWp}  & 2.6 & 2.1 & 1.8 & 1.7 \\ \cline{2-6}
 & \textbf{\textgreater{}100 kWp}  & 1.7 & 1.4 & 1.2 & 1.1 \\ \hline
\end{tabular}
\end{table}

\newpage
\clearpage
\section{Input data of a low, baseline and high cost scenario for battery units}
\label{Apdx:App-futurecosts_battery}

\begin{table}[htbp]
\caption{Baseline battery cost scenario for 2020-2050 \cite{schmidt2019projecting}.}
\label{tab:cost_scen_ref_bat}
\centering
\scriptsize
\begin{tabular}{|l|c|c|c|c|c|c|c|}
\hline
 & \textbf{2020} & \textbf{2025} & \textbf{2030} & \textbf{2035} & \textbf{2040} & \textbf{2045} & \textbf{2050} \\ \hline
\begin{tabular}[c]{@{}l@{}}\textbf{Investment cost} \\ \textbf{(energy-related)} \\ \textbf{\euro/kWh}\end{tabular} & 377 & 233 & 158 & 123 & 110 & 103 & 96 \\ \hline
\begin{tabular}[c]{@{}l@{}}\textbf{Investment cost}\\ \textbf{(power-related)} \\ \textbf{\euro/kW}\end{tabular} & 319 & 197 & 133 & 104 & 93 & 87 & 81 \\ \hline
\begin{tabular}[c]{@{}l@{}}\textbf{Operation cost}\\ \textbf{(energy-related)} \\ \textbf{\euro/MWh}\end{tabular} & 1.41 & 0.87 & 0.59 & 0.46 & 0.41 & 0.38 & 0.36 \\ \hline
\begin{tabular}[c]{@{}l@{}}\textbf{Operation cost}\\ \textbf{(power-related)} \\ \textbf{\euro/kW-year}\end{tabular}
& 4.70 & 2.91 & 1.97 & 1.54 & 1.37 & 1.28 & 1.20 \\ \hline
\end{tabular}
\end{table}

\begin{table}[htbp]
\caption{High battery cost scenario for 2020-2050 \cite{schmidt2019projecting}.}
\label{tab:cost_scen_high_bat}
\centering
\scriptsize
\begin{tabular}{|l|c|c|c|c|c|c|c|c|}
\hline
 & \textbf{2020} & \textbf{2025} & \textbf{2030} & \textbf{2035} & \textbf{2040} & \textbf{2045} & \textbf{2050} \\ \hline
\begin{tabular}[c]{@{}l@{}}\textbf{Investment cost} \\ \textbf{(energy-related)} \\ \textbf{\euro/kWh}\end{tabular} & 459 & 329 & 247 & 206 & 178 & 171 & 158 \\ \hline
\begin{tabular}[c]{@{}l@{}}\textbf{Investment cost}\\ \textbf{(power-related)} \\ \textbf{\euro/kW}\end{tabular} & 388 & 278 & 209 & 174 & 151 & 145 & 133 \\ \hline
\begin{tabular}[c]{@{}l@{}}\textbf{Operation cost}\\ \textbf{(energy-related)} \\ \textbf{\euro/MWh}\end{tabular} & 1.72 & 1.23 & 0.92 & 0.77 & 0.67 & 0.64 & 0.59 \\ \hline
\begin{tabular}[c]{@{}l@{}}\textbf{Operation cost}\\ \textbf{(power-related)} \\ \textbf{\euro/kW-year}\end{tabular}
& 5.73 & 4.10 & 3.08 & 2.56 & 2.22 & 2.14 & 1.97 \\ \hline
\end{tabular}
\end{table}

\begin{table}[htbp]
\caption{Low battery cost scenario for 2020-2050 \cite{schmidt2019projecting}.}
\label{tab:cost_scen_low_bat}
\centering
\scriptsize
\begin{tabular}{|l|c|c|c|c|c|c|c|c|}
\hline
 & \textbf{2020} & \textbf{2025} & \textbf{2030} & \textbf{2035} & \textbf{2040} & \textbf{2045} & \textbf{2050} \\ \hline
\begin{tabular}[c]{@{}l@{}}\textbf{Investment cost} \\ \textbf{(energy-related)} \\ \textbf{\euro/kWh}\end{tabular} & 295 & 137 & 69 & 41 & 41 & 34 & 34 \\ \hline
\begin{tabular}[c]{@{}l@{}}\textbf{Investment cost}\\ \textbf{(power-related)} \\ \textbf{\euro/kW}\end{tabular} & 249 & 116 & 58 & 35 & 35 & 29 & 29 \\ \hline
\begin{tabular}[c]{@{}l@{}}\textbf{Operation cost}\\ \textbf{(energy-related)} \\ \textbf{\euro/MWh}\end{tabular} & 1.10 & 0.51 & 0.26 & 0.15 & 0.15 & 0.13 & 0.13 \\ \hline
\begin{tabular}[c]{@{}l@{}}\textbf{Operation cost}\\ \textbf{(power-related)} \\ \textbf{\euro/kW-year}\end{tabular}
& 3.68 & 1.71 & 0.85 & 0.51 & 0.51 & 0.43 & 0.43 \\ \hline
\end{tabular}
\end{table}

\newpage
\clearpage
\section{Assumed DSO injection tariff by canton }
\label{Apdx:App-PVinjectiontariff}

\begin{table}[htbp]
\renewcommand{\arraystretch}{1.1}
\centering
\caption[DSO injection tariff for PV]{The \gls{dso} injection tariff in cent/kWh for \gls{pv} is estimated for each Swiss Canton \cite{pvtarif}.}
\label{tab:injectiontariff}
\begin{tabular}{|c|c|c|c|c|c|c|c|c|}
\hline
\textbf{Index} & \textbf{Canton} & \textbf{2020} & \textbf{2025} & \textbf{2030} & \textbf{2035} & \textbf{2040} & \textbf{2045} & \textbf{2050} \\ \hline
1 & ZH & 6.6 & 5.6 & 6.1 & 6.5 & 7.0 & 7.6 & 8.2 \\ \hline
2 & BE & 6.9 & 5.6 & 6.1 & 6.5 & 7.0 & 7.6 & 8.2 \\ \hline
3 & LU & 7.3 & 5.6 & 6.1 & 6.5 & 7.0 & 7.6 & 8.2 \\ \hline
4 & UR & 9.0 & 5.6 & 6.1 & 6.5 & 7.0 & 7.6 & 8.2 \\ \hline
5 & SZ & 7.0 & 5.6 & 6.1 & 6.5 & 7.0 & 7.6 & 8.2 \\ \hline
6 & OW & 10.0 & 5.9 & 6.1 & 6.5 & 7.0 & 7.6 & 8.2 \\ \hline
7 & NW & 5.9 & 5.6 & 6.1 & 6.5 & 7.0 & 7.6 & 8.2 \\ \hline
8 & GL & 6.8 & 5.6 & 6.1 & 6.5 & 7.0 & 7.6 & 8.2 \\ \hline
9 & ZG & 11.2 & 6.6 & 6.1 & 6.5 & 7.0 & 7.6 & 8.2 \\ \hline
10 & FR & 8.5 & 5.6 & 6.1 & 6.5 & 7.0 & 7.6 & 8.2 \\ \hline
11 & SO & 8.7 & 5.6 & 6.1 & 6.5 & 7.0 & 7.6 & 8.2 \\ \hline
12 & BS & 11.8 & 7.0 & 6.1 & 6.5 & 7.0 & 7.6 & 8.2 \\ \hline
13 & BL & 9.1 & 5.6 & 6.1 & 6.5 & 7.0 & 7.6 & 8.2 \\ \hline
14 & SH & 7.3 & 5.6 & 6.1 & 6.5 & 7.0 & 7.6 & 8.2 \\ \hline
15 & AR & 5.7 & 5.6 & 6.1 & 6.5 & 7.0 & 7.6 & 8.2 \\ \hline
16 & AI & 9.1 & 5.6 & 6.1 & 6.5 & 7.0 & 7.6 & 8.2 \\ \hline
17 & SG & 8.2 & 5.6 & 6.1 & 6.5 & 7.0 & 7.6 & 8.2 \\ \hline
18 & GR & 9.1 & 5.6 & 6.1 & 6.5 & 7.0 & 7.6 & 8.2 \\ \hline
19 & AG & 6.2 & 5.6 & 6.1 & 6.5 & 7.0 & 7.6 & 8.2 \\ \hline
20 & TG & 7.3 & 5.6 & 6.1 & 6.5 & 7.0 & 7.6 & 8.2 \\ \hline
21 & TI & 8.2 & 5.6 & 6.1 & 6.5 & 7.0 & 7.6 & 8.2 \\ \hline
22 & VD & 7.4 & 5.6 & 6.1 & 6.5 & 7.0 & 7.6 & 8.2 \\ \hline
23 & VS & 7.0 & 5.6 & 6.1 & 6.5 & 7.0 & 7.6 & 8.2 \\ \hline
24 & NE & 8.5 & 5.6 & 6.1 & 6.5 & 7.0 & 7.6 & 8.2 \\ \hline
25 & GE & 11.1 & 6.6 & 6.1 & 6.5 & 7.0 & 7.6 & 8.2 \\ \hline
26 & JU & 6.9 & 5.6 & 6.1 & 6.5 & 7.0 & 7.6 & 8.2 \\ \hline
\end{tabular}
\end{table}
\newpage
\clearpage
\section{Assumed retail electricity tariff by canton }
\label{Apdx:App-elretailtariff}

\begin{table}[htbp]
\renewcommand{\arraystretch}{1.1}
\centering
\begin{threeparttable}
\scriptsize
\caption[Base retail electricity tariff 2020.]{The base retail electricity tariff in cent/kWh estimated for each Swiss Canton and consumption group in 2020 \cite{web_elcom}.}
\label{tab:elretailtariff}
\begin{tabular}{|c|c|c|c|c|c|c|c|c|c|c|c|c|}
\hline
\textbf{Canton} & \textbf{L1} & \textbf{L2} & \textbf{L3} & \textbf{L4} & \textbf{L5} & \textbf{L6} & \textbf{L7} & \textbf{L8} & \textbf{L9} & \textbf{L10} & \textbf{L11}\\ \hline
ZH & 20.5 & 18.3 & 16.7 & 16.7 & 15.8 & 15.8 & 15.5 & 14.3 & 12.3 & 16.3 & 14.3 \\ \hline
BE & 27.7 & 25.1 & 23.1 & 23.1 & 22.1 & 22.1 & 22.2 & 19.7 & 17.6 & 22.3 & 19.9 \\ \hline
LU & 23.0 & 22.7 & 21.0 & 21.0 & 20.7 & 20.7 & 22.1 & 17.8 & 14.9 & 21.0 & 17.0 \\ \hline
UR & 28.4 & 25.5 & 23.3 & 23.3 & 22.2 & 22.2 & 22.2 & 19.2 & 15.8 & 20.9 & 16.7 \\ \hline
SZ & 23.9 & 21.7 & 20.1 & 20.1 & 19.2 & 19.2 & 18.7 & 16.9 & 15.0 & 19.2 & 16.7 \\ \hline
OW & 26.9 & 24.0 & 22.1 & 22.1 & 21.0 & 21.0 & 20.8 & 19.1 & 16.6 & 19.9 & 17.7 \\ \hline
NW & 24.2 & 21.4 & 19.9 & 19.9 & 18.8 & 18.8 & 18.1 & 17.0 & 15.6 & 17.5 & 16.1 \\ \hline
GL & 27.7 & 25.1 & 22.3 & 22.3 & 20.1 & 20.1 & 18.5 & 17.1 & 15.1 & 20.9 & 19.5 \\ \hline
ZG & 21.0 & 20.1 & 18.2 & 18.2 & 17.4 & 17.4 & 17.7 & 14.9 & 12.8 & 18.1 & 15.3 \\ \hline
FR & 24.5 & 22.2 & 20.1 & 20.1 & 19.3 & 19.3 & 19.8 & 17.3 & 14.0 & 19.9 & 19.1 \\ \hline
SO & 26.1 & 23.2 & 21.6 & 21.6 & 20.7 & 20.7 & 20.6 & 18.6 & 16.2 & 21.0 & 18.8 \\ \hline
BS & 27.4 & 27.2 & 25.8 & 25.8 & 25.9 & 25.9 & 27.3 & 23.1 & 21.0 & 29.6 & 25.6 \\ \hline
BL & 25.4 & 23.0 & 21.5 & 21.5 & 20.7 & 20.7 & 20.4 & 17.1 & 17.2 & 20.7 & 18.8 \\ \hline
SH & 24.8 & 22.1 & 20.6 & 20.6 & 19.6 & 19.6 & 19.1 & 16.4 & 15.3 & 19.3 & 16.7 \\ \hline
AR & 21.0 & 19.3 & 17.8 & 17.8 & 16.6 & 16.6 & 16.0 & 14.2 & 13.1 & 16.3 & 13.5 \\ \hline
AI & 22.2 & 19.2 & 17.6 & 17.6 & 16.5 & 16.5 & 15.9 & 14.3 & 13.0 & 16.2 & 14.1 \\ \hline
SG & 23.2 & 20.5 & 18.8 & 18.8 & 17.7 & 17.7 & 17.0 & 15.5 & 14.1 & 17.8 & 15.2 \\ \hline
GR & 24.9 & 22.2 & 21.1 & 21.1 & 20.4 & 20.4 & 20.4 & 18.8 & 16.6 & 19.9 & 19.6 \\ \hline
AG & 26.8 & 20.4 & 18.7 & 18.7 & 17.6 & 17.6 & 17.0 & 15.5 & 13.3 & 17.1 & 16.2 \\ \hline
TG & 23.2 & 20.6 & 19.0 & 19.0 & 17.9 & 17.9 & 17.3 & 159 & 14.3 & 18.6 & 16.5 \\ \hline
TI & 21.4 & 20.2 & 18.9 & 18.9 & 18.6 & 18.6 & 18.6 & 17.0 & 15.6 & 18.8 & 19.2 \\ \hline
VD & 24.1 & 22.6 & 20.9 & 20.9 & 20.2 & 20.2 & 20.5 & 183 & 15.7 & 20.3 & 17.7 \\ \hline
VS & 20.3 & 18.3 & 17.3 & 17.3 & 18.6 & 18.6 & 15.4 & 16.0 & 13.3 & 16.8 & 15.2 \\ \hline
NE & 25.4 & 23.8 & 214 & 21.4 & 20.1 & 20.1 & 20.2 & 18.0 & 14.9 & 21.1 & 18.8 \\ \hline
GE & 20.5 & 20.2 & 19.1 & 19.1 & 19.0 & 19.0 & 20.0 & 18.1 & 16.2 & 20.7 & 19.5 \\ \hline
JU & 32.2 & 28.6 & 26.3 & 26.3 & 25.3 & 25.3 & 25.6 & 21.1 & 17.6 & 25.5 & 21.8 \\ \hline
\end{tabular}
    \begin{tablenotes}
      \scriptsize
      \item Note: the low retail electricity tariff for off-peak hours and the high retail electricity tariff for peak hours are assumed to be 71\% and 107\% of the base tariff.
      \end{tablenotes}
\end{threeparttable}
\end{table}

\begin{table}[htbp]
\renewcommand{\arraystretch}{1.2}
\caption{Information of electricity consumption categories \cite{web_elcom}.}
\label{tab:consumptionCategory}
\begin{tabular}{|c|c|c|}
\hline
\textbf{Category} & \textbf{Annual electricity consumption} & \textbf{Electricity tariff category} \\ \hline
\textbf{L1} & 0-1'600 kWh & H1 \\ \hline
\textbf{L2} & 1'600-2'500 kWh & H2 \\ \hline
\textbf{L3} & 2'500-3'500 kWh & H2, H3 \\ \hline
\textbf{L4} & 3'500-4'500 kWh & H2, H3 \\ \hline
\textbf{L5} & 4'500-5'500 kWh & H3, H4 \\ \hline
\textbf{L6} & 5'500-7'500 kWh & H3, H4 \\ \hline
\textbf{L7} & 7'500-13'000 kWh & H8 \\ \hline
\textbf{L8} & 13'000-25'000 kWh & H7 \\ \hline
\textbf{L9} & 25'000-30'000 kWh & H6 \\ \hline
\textbf{L10} & 30'000-150'000 kWh & C2 \\ \hline
\textbf{L11} & >150'000 kWh & C3 \\ \hline
\end{tabular}
\end{table}

\newpage
\clearpage

\end{appendices}


\end{document}